\documentclass[pre, twocolumn, showpacs, amsmath,superscriptaddress]{revtex4}
\usepackage{dcolumn}
\usepackage{bm}
\usepackage{graphicx}
\usepackage{color}
\usepackage{xcolor}
\usepackage{amsmath}
\DeclareGraphicsExtensions{. jpg,. pdf, . mps, . png, . eps, . ps, . EPS}

\newcommand{\mediaT}[1]{\left\langle #1 \right\rangle}

\def\bc{\begin{center}} 
\def\ec{\end{center}}
\def\bea{\begin{eqnarray}}
\def\eea{\end{eqnarray}}
\newcommand{\avg}[1]{\langle{#1}\rangle}

\newcommand{\comment}[1]{{#1}} 

\begin{document}

\title{Statistical mechanics of random geometric graphs: \\ 
Geometry-induced first order phase transition}

\author{Massimo Ostilli}
\affiliation{
Departamento de Fisica, Universidade Federal de Santa Catarina, Florianopolis 88040-900, Santa Catarina, Brazil} 
\affiliation{Dipartimento di Fisica, Universit\`a di Roma ‘La Sapienza’, Piazzale Aldo Moro 2, Roma I-00185, Italy}
\author{Ginestra Bianconi}
\affiliation{Queen Mary University of London, London, E1 4NS, United Kingdom}

\begin{abstract}
Random geometric graphs (RGG) can be formalized as 
hidden-variables models where the hidden variables are
the coordinates of the nodes.  Here we develop a general approach 
to extract  the typical configurations of a generic hidden-variables
model and apply the resulting equations to RGG. 
For any RGG, defined through a rigid or a soft geometric rule, 
the method reduces to  a non trivial satisfaction problem:
Given $N$ nodes, a domain $\mathcal{D}$, and a desired average connectivity $\avg{k}$,
find - if any - the distribution of nodes having support in $\mathcal{D}$ and
average connectivity $\avg{k}$. We find out that, in the thermodynamic limit,
nodes are either uniformly distributed or highly condensed in a small region, the two regimes
being separated by a first order phase transition characterized by a $\mathop{O}(N)$ jump of $\avg{k}$.
Other intermediate values of $\avg{k}$ correspond to very rare graph realizations.
The phase transition is observed as a function of a parameter $a\in[0,1]$ that tunes the underlying geometry.
In particular, $a=1$ indicates a rigid geometry where only close nodes are connected, 
while  $a=0$ indicates a rigid anti-geometry where only distant nodes are connected. 
Consistently, when $a=1/2$ there is no geometry and no phase transition. 
After discussing the numerical analysis, we provide a combinatorial argument to fully explain the mechanism inducing this phase transition
and recognize it as an easy-hard-easy transition.
Our result shows that, in general, ad hoc optimized networks can hardly be designed, unless to rely to
specific heterogeneous constructions, not necessarily scale free.  
\end{abstract}

\pacs{89.75.Hc, 05.70.Fh,  05.70.Ce, 02.70.-c}

\maketitle

\section{Introduction}
In the last decades, statistical mechanics has seen an outstanding progress  
as a powerful mathematical tool, physically oriented, for analyzing complex systems of the most
diverse nature. A special impulse was later given to the interplay between concepts
close to the realm of statistical physics, mostly within the spin glass theory \cite{Parisi}, and the mathematical theory of complexity
for optimization problems \cite{Zecchina,Zdeborova}.
In more recent years, the developments in complex network theory and its numerous applications 
have made even stronger the importance of statistical mechanics as a very interdisciplinary mathematical tool
\cite{DMAB,Review,SND,Bianconi1,Estrada,Bianconi3}.
In network theory, elements and interactions are simplified as nodes and links between them. 
Several models for growing or static networks have been
proposed and turned out able to reproduce the main universal features observed in real networks.
One of the most promising frameworks to analyze networks is provided by hidden variables models 
\cite{CaldarelliHidden,BogunaHidden,NewmanHidden,SatorrasHidden}. 
The general idea behind hidden variable models
relies on the possibility that each node is attributed a value $h$ that expresses its propensity to be connected to another node with hidden variable $h'$ via
a formula that relates $h$ with $h'$.
It turns out that this scheme is both very general and very powerful in the description of many diverse networks, especially with respect to
the possibility of having analytic results.

In this paper, we use statistical mechanics to face and solve two concatenated problems in hidden variable models.
We first develop a general approach to extrapolate the typical configurations of a generic hidden-variables
model, \textit{i.e.}, we use statistical mechanics to derive, in the thermodynamic limit, the
equations that describe, in the space of the hidden variables $h$, the node distribution ${{\nu}}_h$ corresponding to the most important graph realizations 
(the others being exponentially rare in the thermodynamic limit). Then, we apply these general equations to
random geometric graphs (RGG) \cite{Meester,Dall,Penrose} where the $h$'s represent vector positions. 
Unlike random or common complex network models, in RGG nodes are connected according 
to an underlying geometry (Euclidean or not). In particular, in classical RGG  two nodes are connected 
if their distance is at most equal to a given threshold. 
RGG are important in many theoretical and practical aspects, but despite that, the exact analysis of RGG
has been mostly confined to the Euclidean $d=1$ and $d=2$ dimensional cases, 
and mainly focused on the percolation problem.
Notice that, when $d>1$, RGG are not tree-like networks.
In fact, they have dense loops, \textit{i.e.}, the probability that a random walk of $l$ links
is closed remains finite for any $l\geq 3$, even in the infinite size limit (for $l=3$ this amounts to a finite clustering coefficient).  
This fact avoids using the same probabilistic techniques, 
like the Belief-Propagation (or Bethe-Peierls) methods, which are exact only on tree-like graphs. 
It is worth mentioning here the Kasteleyn and Fortuin approach to the percolation problem \cite{Kasteleyn}, which maps
the counting of connected clusters toward the mean-field Potts model;
an outstanding example of how statistical mechanics can be used to solve combinatorial problems (see also \cite{Zecchina}).
However, when we are dealing with RGG, due to the presence of dense loops, the Kasteleyn and Fortuin approach would require
to solve a non mean-field
Potts model, an analytically unfeasible task.

\begin{figure}[htbp]
{\includegraphics[width=3.2in]{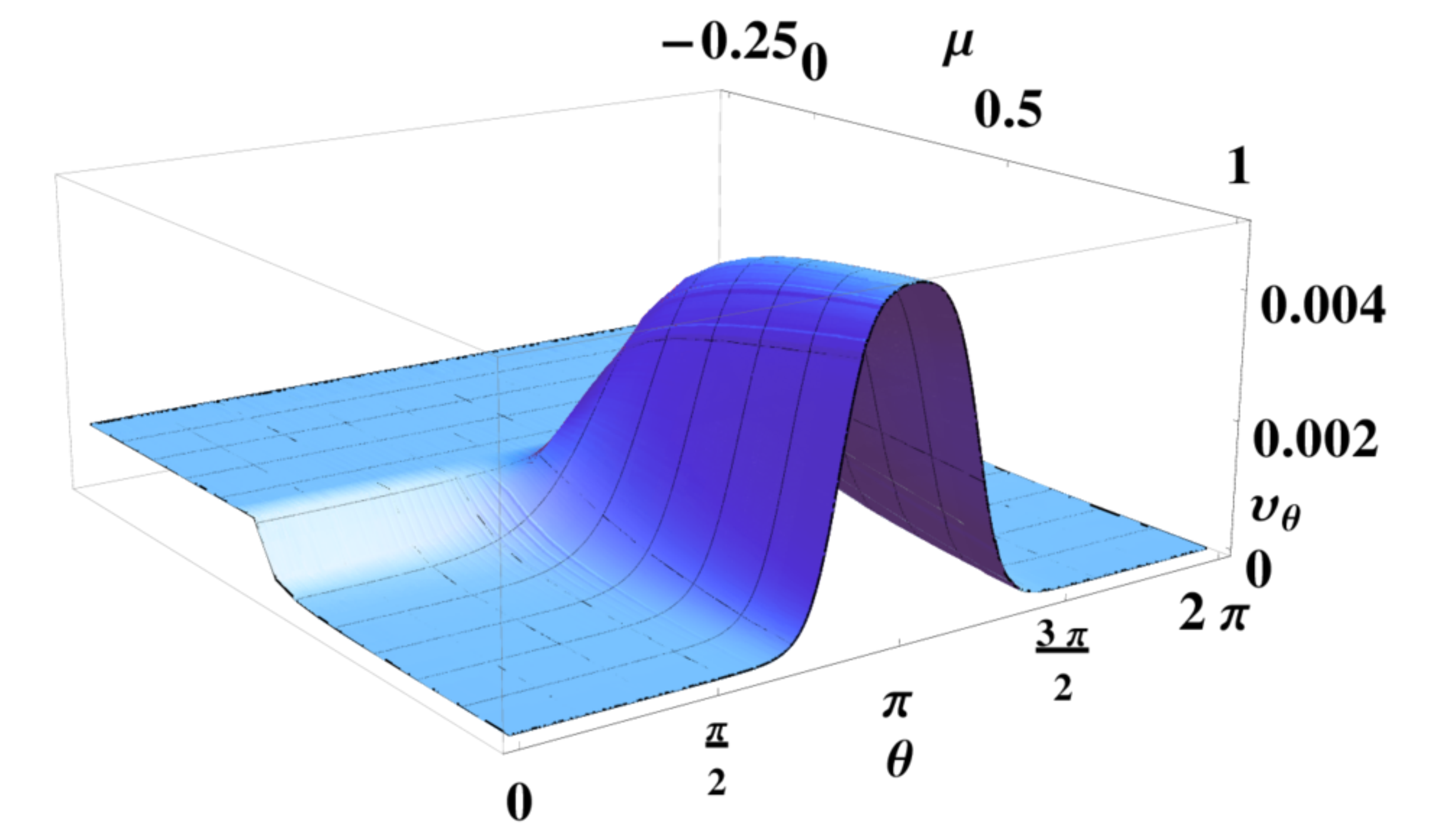}}
{\includegraphics[width=3.2in]{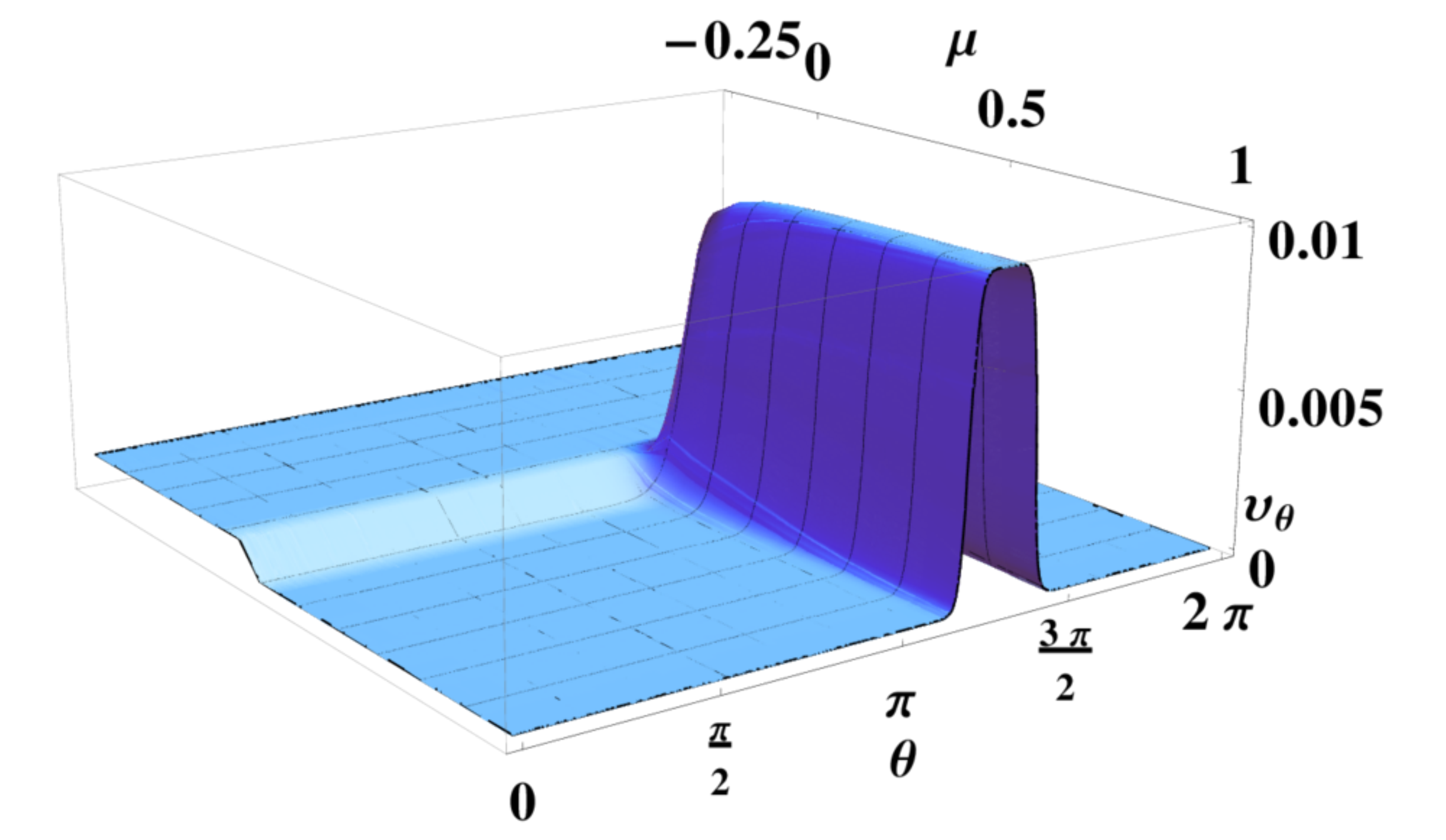}}
{\includegraphics[width=3.2in]{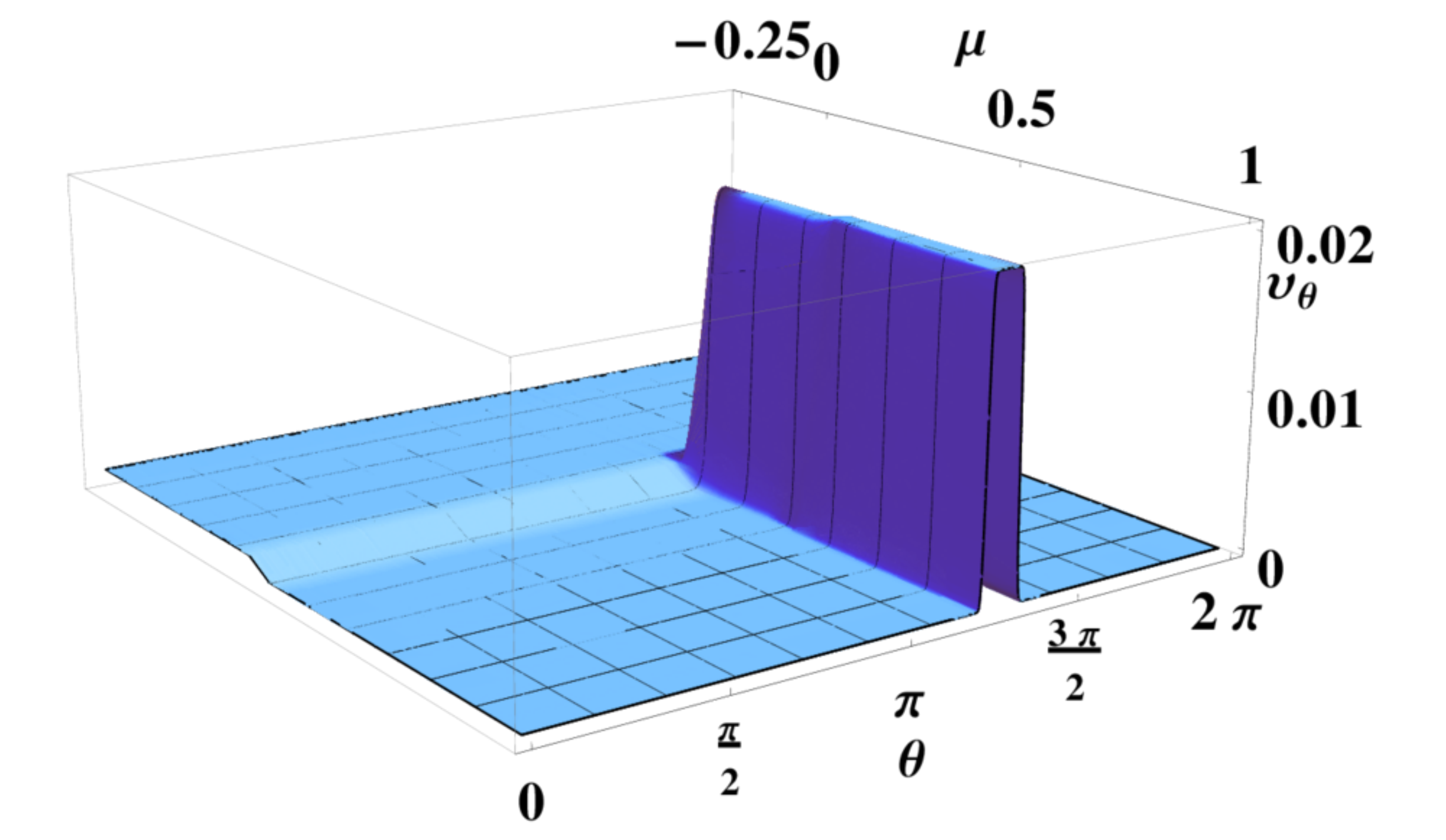}}
\caption{\comment{(Color online) Typical node-distribution (associated to the typical graphs) for a 
RGG defined over the circle
according to a simple geometric rule (\ref{rule})-(\ref{d0}) with a rigid constrain ($a=1$). 
The plot gives the distribution ${{\nu}}_\theta$ as a function of the angle $\theta$ for each value of the chemical potential $\mu$.
As the number of nodes $N$ increases, there emerges a separation of two phases in correspondence of a critical value of the
chemical potential $\mu_c\simeq 0.25$.
For $\mu<\mu_c$ the nodes tend to be uniformly distributed and have a minimal average connectivity $\avg{k}=\mathop{O}(1)$, 
whereas, for $\mu>\mu_c$, 
a condensation of nodes takes place and tend to form a fully connected graph with $\avg{k}=\mathop{O}(N)$.
Top to bottom: $N=10,~20,~40$.}
\label{fig0}
}
\end{figure}

In this paper we adopt a different approach. 
We do not aim to analyzing the percolation problem in RGG, nor to have a full descriptive solution of RGG
(which would allow us to know all the graph metrics).
Rather, we look for reduced descriptive, yet exact, solutions.  
By using the hidden variable framework, and our general result for the typical configurations, we face a practical 
satisfaction problem:
Given $N$ nodes, a domain $\mathcal{D}$, and a desired average connectivity $\avg{k}$,
find - if any - the distribution of nodes having support in $\mathcal{D}$ and
average connectivity $\avg{k}$. Notice that this is a typical problem in ad hoc wireless optimized networks \cite{ADHOC}:
Embedded in a space with $d=2$ or $d=3$ dimensions, one has $N$ mobile devices that must be connected, 
in average, with a certain number of neighbors
at a minimal cost. Furthermore, it might also be required that the whole structure be a connected graph.
However, we find out that, in the thermodynamic limit,
nodes can only be either uniformly distributed or highly condensed in a small region, the two regimes
being separated by a first order phase transition characterized by a $\mathop{O}(N)$ jump of $\avg{k}$ 
(see Fig.\ref{fig0} for an example whose details will be discussed later).
Other intermediate values of $\avg{k}$ correspond only to very rare graph realizations.
The phase transition is triggered by the underlying geometry whose strength can be tuned by a parameter $a\in[0,1]$,
with $a=1$ for rigid geometry (only close nodes are connected) and $a=0$ for rigid anti-geometry (only distant nodes are connected). 
Consistently, when $a=1/2$ there is no geometry and no phase transition. 
After discussing the numerical solution for RGG in $d=1$ dimensions, 
we provide a combinatorial argument to fully explain the mechanism inducing this phase transition in general RGG.
In turn, the combinatorial argument leads us to recognize the phase transition as an easy-hard-ease transition, the hard solutions
of the problem being located in between the uniform and condensed regimes, \textit{i.e.}, at the critical point. 
Furthermore, we show that the hard solutions corresponding to connected graphs are
networks quite heterogeneous.
Our result shows that, in general, ad hoc optimized networks can hardly be designed, unless to
rely to strongly heterogeneous constructions, not necessarily power law.  

\section{Hidden-Variables Models}
Given $N$ nodes, hidden variable models are defined in the following way:
\textit{i)}
to each node we associate a hidden variable $h$ drawn from
a given probability density function (PDF) $\rho(h)$; 
\textit{ii)} between any pair of nodes, we assign a link according
to a given set of probabilities $p(h,h')$, where $h$ and $h'$ are the hidden variables associated to the two nodes.
The probabilities $p(h,h')$ can be any function of the $h$'s, the only requirement being that $0 \leq p(h,h')\leq 1$.
The hidden variables can be real numbers or also vectors. Many networks can be embedded in a hidden-variables scheme.
For example, in RGG the $h$' s are vectors representing the positions of the nodes,
and the $p(h,h')$'s are defined in terms of geometric rules (non deterministic if $0<p(h,h')<1$, or deterministic if $p(h,h')$ takes only the values 0 or 1).
Notice that, by construction, the steps \textit{i)}-\textit{ii)} generate only simple graphs (\textit{i.e.}, without
multiple edges or loop-edges).

Particular attention has been paid to the ``configuration model''. In this case
$p(h,h')$ has the Fermi-Dirac form (or similar generalizations)~
\begin{eqnarray}
\label{CM}
&& p(h,h')= \left(1+\frac{k_s^2}{hh'}\right)^{-1}, \quad
k_s=\sqrt{N\bar{k}},
\end{eqnarray}
where, for large $N$, $\bar{k}$ coincides with the average degree. 
In general, the actual degree $k$ of the nodes of the network realized with the above scheme
are distributed according to $\rho$ and average degree equal to $\bar{k}$.
In particular, if we choose the following PDF having support in $[h_{\mathrm{min}},h_{\mathrm{max}}]$
\begin{eqnarray}
\label{rho}
\rho(h)=a~h^{-\gamma}, \quad h_{\mathrm{max}} \geq h \geq h_{\mathrm{min}}>0,
\end{eqnarray}
with $\gamma>2$,
the degree-distribution $p(k)$ of the resulting network will be a power law with exponent $\gamma$.

For what follows, is important to note that Eqs. (\ref{CM}) and (\ref{rho}) represent a very particular hidden-variables model.
\comment{In fact, for such a specific example we have full control of the average number of links via the parameter $\overline{k}$ inside the
function $p(h,h')$. Notice that, given any couple $p(h,h')$ and $\rho(h)$, the hidden variable scheme defined through the steps \textit{i)-ii)} determines
an ensemble of graphs whose properties, like average connectivity, degree distribution, density of triangles etc..., are all mathematically encoded in $p(h,h')$ and $\rho(h)$.
However, only in a few cases the relation between these ensemble properties and the parameters entering the functions $p(h,h')$ and $\rho(h)$ turn out to be relatively simple.
In general, given $p(h,h')$ and $\rho(h)$, we are not able to control directly the ensemble properties via these parameters, not even the average number of links.
In fact, this is the case for RGG. Therefore, for the sake of generality,
we will allow for the presence of a chemical potential $\mu$ to be adjusted toward a suitable value in order to have 
a desired average number of links.}

\newpage
\section{Typical Configurations in Hidden-Variables Models}

\subsection{Motivation, formulation of the problem and synthesis of the method}
Let be given a hidden-variables model through the functions $p(h,h')$ and $\rho(h)$.
There are two kinds of network ensembles that can be generated. 
Inspired by statistical mechanics, we call these two ensembles: canonical and gran canonical.

\textit{Canonical Ensemble.}
If we make a \textit{una tantum} sampling of the $h$'s and 
these are kept fixed, we are dealing with an ensemble of graphs where the $h$'s play the role
of parameters, while the random variables are the elements of the adjacency matrix $\{a_{i,j}\}$,
each taking the values $1$ or $0$ for the presence or absence of a link, respectively. 
Examples include networks whose nodes have fixed expected degrees and, more in general,
networks that can be described as exponential random graphs \cite{Newman}. 
One peculiar feature of the Canonical Ensemble is that 
the relative fluctuations of graph-observables that are extensive in the system size tend to 0 for $N\to \infty$.
This fact allows to extrapolate many characteristics of the ensemble 
which we can define as ``typical'' \cite{Bianconi1,Bianconi3}.

\textit{Gran Canonical Ensemble.}
This is the kind of ensemble we will deal with.
In this case the $h$'s are random variables of the ensemble together with the $\{a_{i,j}\}$.
An important example are RGG where the $h$'s represent the coordinates of the nodes.
Another example is provided by gradient networks, where the $h$'s determine the flow directions \cite{Gradient}. 
However,  the Gran Canonical Ensemble is not just an enlargement of the Canonical one; it has a quite different nature. 
Unlike the Canonical Ensemble, in the Gran Canonical Ensemble, in general, fluctuations of extensive quantities 
cannot be neglected. 
In particular, it has been shown that, in the scale-free model (\ref{CM})-(\ref{rho}), 
the number of links and the entropy are the only observables 
whose relative fluctuations can be neglected when $N\to\infty$ \cite{NSA,NSA1,Bianconi4}.
In order to calculate the ensemble averages, $\avg{\cdot}$, 
one has to evaluate a generalized partition function,
or functional generator, $Z(\{x_{i,j}\};\{y_i\})$, 
where the $x_{i,j}$ and the $y_i$ are external fields conjugated to the variables $a_{i,j}$ and $h_i$, respectively.
A manifestation of the large fluctuations of the Gran Canonical Ensemble 
is that $Z(\{x_{i,j}\};\{y_i\})$ cannot be calculated via
saddle point techniques (in Appendix \ref{appA} we show this in detail).
However, under certain conditions, we can still find  
some typical features of the ensemble, exactly. 
In fact, if we limit the description of a graph to its hidden-variables sequence, 
such a plan is mathematically possible.
To this aim, we analyze the generating function $Z(\mu)$ 
where $\mu$ is conjugated to the number of links $\sum_{i<j}a_{i,j}$ 
and plays the role of the chemical potential of the ensemble. 
Notice that, in general, there is a one-to-one correspondence between $\mu$   
and the average connectivity $\avg{k}$: $\mu=\mu(\avg{k})$.
We then perform the change of variables $\{h_i\}\to \{N_h\}$,
where $N_h=\sum_i \delta(h_i,h)$. 
With this change of variables, for any $\mu\in I$, where $I$ is a finite interval, 
the integrand in $Z(\mu)$ is highly concentrated around its maximum, 
and $Z(\mu)\propto\exp (-F(\{N_h^*\}))$, where $F(\{N_h^*\})$ plays the
role of a free energy with respect to the variables $\{N_h\}$, 
and the $N_h^*=N {{\nu}}_h^*$ are solution of certain saddle point Eqs.. 
As we will show, given $N$, $\bm{p}=\{p_{h,h'}\}$, $\bm{\rho}=\{\rho_h\}$, and a desired average connectivity $\avg{k}$, 
the saddle point Eqs. for ${{\nu}}_h$ and $\mu$ lead to a crucial duality relation between the
ensemble characterized by $\bm{p}=\{p_{h,h'}\}$, $\bm{\rho}=\{\rho_h\}$ and $\mu=\mu(\avg{k})$, and an ensemble characterized 
by ${\bm{q}}_\mu=\{q_{h,h'}\}$, ${\bm{\nu}}_\mu=\{\nu_h^*\}$,  
and $\mu=0$: 
\begin{eqnarray}
\label{Dual0} 
Ensemble\left\{\bm{p},\bm{\rho},\mu\right\}=Ensemble\left\{{\bm{q}}_\mu,{\bm{\nu}}_\mu,0\right\}, 
\end{eqnarray}
where ${\bm{q}}_\mu$ are a proper modification of the probabilities $\bm{p}$. 
To simplify the notation we shall often omit the suffix $\mu$ in ${\bm{\nu}}_\mu$ and ${\bm{q}}_\mu$.
The duality relation (\ref{Dual0}) allows to draw, in principle, networks with arbitrary $\avg{k}$ without changing~$N$.
In particular, as we shall see, for rigid RGG we have ${\bm{q}}_\mu={\bm{p}}$, but ${\bm{\nu}}_\mu\neq \bm{\rho}$.
In the following of this Section we derive the saddle point Eqs. that lead to Eq. (\ref{Dual0}) and the
relation between $\avg{k}$ and $\mu$.

\subsection{Analysis}
Let be given a hidden-variables model through the functions $p(h,h')$ and $\rho(h)$.
The conditional probability to realize a graph $\bm{G}=\{a_{i,j}\}$ with frozen hidden variables $\{h_i\}$ is
\begin{eqnarray}
\label{Cond}
&& P(\bm{G}|\{h_i\})=\prod_{i<j}\left[p(h_i,h_j)\right]^{a_{i,j}}\left[1-p(h_i,h_j)\right]^{1-a_{i,j}},
\end{eqnarray}
while the (unconditioned) probability to have $\bm{G}$ is
\begin{eqnarray}
\label{Full1}
&& P(\bm{G})=\int \prod_{i}dh_i \rho(h_i)\prod_{i<j}\left[p(h_i,h_j)\right]^{a_{i,j}}\nonumber \\
&& \times \left[1-p(h_i,h_j)\right]^{1-a_{i,j}}.
\end{eqnarray}

The generating function we want to calculate is
\begin{eqnarray}
\label{Z}
&& Z(\mu)= \int \prod_i dh_i \rho(h_i) \sum_{\{a_{i,j}\}}P(\bm{G}|\{h_i\}) e^{\mu \sum_{i<j}a_{i,j}},
\end{eqnarray}
and the average number of links $\mediaT{L}$ is given by
\begin{eqnarray}
\label{L}
\mediaT{L}=\frac{\partial \ln Z(\mu)}{\partial \mu}.
\end{eqnarray}
By performing the sum over the $\{a_{i,j}\}$ we obtain
\begin{eqnarray}
\label{ZA2}
&& Z(\mu)= \int \prod_i dh_i e^{-G(\{h_i\})}, 
\end{eqnarray}   
where 
\begin{eqnarray}
\label{G}
&& -G(\{h_i\})=\sum_i \ln[\rho(h_i)] +\nonumber \\
&& \sum_{i<j}\ln\left[p(h_i,h_j)e^\mu+1-p(h_i,h_j)\right]
\end{eqnarray}   
For the sake of clarity, we will suppose to deal with a discrete distribution for the $h$'s, on a finite set $\mathcal{D}$,
and to stress this, we will write $\rho_h$ instead of $\rho(h)$:
\begin{eqnarray}
\label{rhod}
\sum_{h\in \mathcal{D}} \rho_h=1.
\end{eqnarray}   
Later on we will extrapolate the continuum limit too.
We notice now that $G(\{h_i\}$ depends only on the multiplicities of the $h$'s:
\begin{eqnarray}
\label{G1}
&& -G(\{h_i\})=-\tilde{G}\left(\{N_h\}\right)
\nonumber \\
&&=\sum_h N_h\ln(\rho_h) +\frac{1}{2}\sum_{h,h'}N_hN_{h'}\sigma_{h,h'},
\end{eqnarray}   
where ($\delta(x,y)$ stands for Kronecker delta function)
\begin{eqnarray}
\label{G2}
N_h=\sum_i \delta(h_i,h), 
\end{eqnarray}    
\textit{i.e.}, given any ensemble realization $\{h_i\}$, for any $h\in \mathcal{D}$, $N_h$ provides the number of nodes with hidden variable $h$, and
\begin{eqnarray}
\label{SQ}
\sigma_{h,h'}=\ln\left[p(h,h')e^\mu+1-p(h,h')\right]
\end{eqnarray}    
By using Eqs. (\ref{G}-\ref{SQ}), $Z(\mu)$ becomes
\begin{eqnarray}
\label{ZA3}
&& Z(\mu)= \sum'_{\{N_h\}}\mathcal{N}\left(\{N_h\}\right) \nonumber \\
&& \times e^{\sum_h N_h\ln(\rho_h) +\frac{1}{2}\sum_{h,h'}N_hN_{h'}\sigma_{h,h'}}
\end{eqnarray}  
where, $\mathcal{N}\left(\{N_h\}\right)$ is the number of ways in which we can arrange 
$N$ nodes such that the multiplicities of the $h$'s are $\{N_h\}$, and $\sum'_{\{N_h\}}$ stands for sum over all the normalized
multiplicities $\sum_h N_h=N$. The number $\mathcal{N}\left(\{N_h\}\right)$ is given by
\begin{eqnarray}
\label{NN}
\mathcal{N}\left(\{N_h\}\right)=\frac{N!}{\prod_h N_h!},
\end{eqnarray}    
therefore, by using the Stirling approximation, Eq. (\ref{ZA3}) becomes
\begin{eqnarray}
\label{ZA4}
&& Z(\mu)= \sum'_{\{N_h\}} e^{-F\left(\{N_h\}\right)},
\end{eqnarray}   
where now
\begin{eqnarray}
\label{FA}
&& -F\left(\{N_h\}\right)=N\ln(N)+\sum_h N_h\left(\ln(\rho_h)-\ln(N_h)\right) \nonumber \\
&& +\frac{1}{2}\sum_{h,h'}N_hN_{h'}\sigma_{h,h'}.
\end{eqnarray}   
We will evaluate the sum in Eq. (\ref{ZA4}) by making use of
\begin{eqnarray}
\label{SS}
&& \sum'_{\{N_h\}} \sim \int' \prod_h dN_h=\int \prod_h dN_h\delta\left(N-\sum_h N_h\right)\nonumber \\
&& =\frac{1}{2\pi}\int \prod_h dN_h \int dx e^{ix\left(N-\sum_h N_h\right)}.   
\end{eqnarray}   
The $N_h$'s are extensive. Let us rewrite them as 
\begin{eqnarray}
\label{NN1}
N_h=N {{\nu}}_h.
\end{eqnarray}
With these notations we have
\begin{eqnarray}
\label{ZA5}
&& Z(\mu)= \sum'_{\{N_h\}} e^{-F\left(\{N_h\}\right)}\nonumber \\
&&\sim \frac{N^{|\mathcal{D}|}}{2\pi}\int \prod_h d{{\nu}}_h \int dx e^{-Nf\left(\{{{\nu}}_h\};x\right)},
\end{eqnarray}   
where $|\mathcal{D}|$ is the cardinality of the set of the $h$'s, $\mathcal{D}$, and 
\begin{eqnarray}
\label{FA1}
&& -f\left(\{{{\nu}}_h\};x\right)=\ln(N)\left(1-\sum_h {{\nu}}_h\right)\nonumber \\
&& +ix\left(1-\sum_h {{\nu}}_h\right) +\sum_h {{\nu}}_h\left(\ln(\rho_h)-\ln({{\nu}}_h)\right) \nonumber \\
&& +\frac{1}{2}N\sum_{h,h'}{{\nu}}_h{{\nu}}_{h'}\sigma_{h,h'}.
\end{eqnarray}   
Of course, in the continuum limit $|\mathcal{D}|\to\infty$, so that $Z(\mu)$ is ill defined, but the averages are all well defined. 
We have 
\begin{eqnarray}
\label{PD1}
&& -\frac{\partial f}{{{\nu}}_h} =-\ln(N) -ix + \ln(\rho_h)-\ln({{\nu}}_h) -1 \nonumber \\
&& +N\sum_{h'}{{\nu}}_{h'}\sigma_{h,h'},
\end{eqnarray}   
and
\begin{eqnarray}
\label{PD2}
-\frac{\partial f}{\partial x} = i\left(1-\sum_h {{\nu}}_h\right).
\end{eqnarray}   
The saddle point Eqs. $\nabla f=0$ imply that 
Eq. (\ref{PD2}) fixes the normalization of the ${{\nu}}_h$'s: $\sum_h {{\nu}}_h=1$, 
so that
\begin{eqnarray}
\label{sp}
{{\nu}}_h=\frac{\rho_he^{N\sum_{h'}{{\nu}}_{h'}\sigma_{h,h'}}}{\sum_{h''}\rho_{h''}e^{N\sum_{h'}{{\nu}}_{h'}\sigma_{h'',h'}}}.
\end{eqnarray}   
These Eqs. remain well defined also in the continuum limit $|\mathcal{D}|\to\infty$, being
\begin{eqnarray}
\label{spC}
{{\nu}}(h)=\frac{\rho(h)e^{N\int dh' {{\nu}}(h')\sigma(h,h')}}{\int dh''\rho(h'')e^{N\int dh'n(h')\sigma(h,h')}}.
\end{eqnarray}   

Once solved, Eqs. (\ref{sp}) (or Eqs. (\ref{spC})), give the saddle point solution ${{\nu}}_h^*$ 
to be plugged into the expression for $Z(\mu)$,
and provide the most likelihood sequences $\{{{\nu}}_h^*\}$ that maximizes $\tilde{P}_\mu\left(\{{{\nu}}_h\}\right)$, 
defined as the probability to pick up at random any graph with given sequence $\{{{\nu}}_h\}$ in the presence of a chemical potential $\mu$:
\begin{eqnarray}
\label{ZA6}
&& Z(\mu) \propto e^{-Nf\left(\{{{\nu}}_h^*\}\right)} \nonumber \\
&& \sim \max_{\{{{\nu}}_h\}} \left\{\mathcal{N}\left(\{{{\nu}}_h\}\right) \prod_h \rho_h^{N_h} P_\mu(\{N_h\})\right\}\nonumber \\
&& \propto \max_{\{{{\nu}}_h\}} \left\{ \tilde{P}_\mu(\{N_h\}) \right\},
\end{eqnarray}   
where (with an abuse of notation, on noticing that $f\left(\{{{\nu}}_h^*\};x\right)$ does not depend on $x$, we write $f\left(\{{{\nu}}_h^*\}\right)$ instead of $f\left(\{{{\nu}}_h^*\};x\right)$)
\begin{eqnarray}
\label{Fsp}
&&-f\left(\{{{\nu}}_h^*\}\right)=\sum_h {{\nu}}_h^*\left(\ln(\rho_h)-\ln({{\nu}}_h^*)\right) \nonumber \\
&&+\frac{1}{2}N\sum_{h,h'}{{\nu}}_h^*{{\nu}}_{h'}^*\sigma_{h,h'}.
\end{eqnarray}   
 
Numerically, we can solve Eqs. (\ref{sp}) only for discrete $h$'s. 
In some cases we can manage also the continuum limit analytically.

\subsection{The role of the chemical potential $\mu$; complete system of Eqs.}
From Eq. (\ref{SQ}) we  see that 
\begin{eqnarray}
\label{SQ1}
\operatorname{sgn}\left(\sigma_{h,h'}\right)=\operatorname{sgn}\left(\mu\right),
\end{eqnarray}    
hence, when $\mu\to 0$,  $\sigma_{h,h'}\to 0$ and, 
as a consequence, Eqs. (\ref{sp}) give ${{\nu}}_h^*\to \rho_h$. 
As we mentioned in Sec. II, in the case of the scale-free model (\ref{CM})-(\ref{rho}),   
we can control the average connectivity $\avg{k}=2\avg{L}/N$ via the parameter $\bar{k}$,
by simply choosing $\bar{k}=\avg{k}$, 
and we do not need to operate with a $\mu\neq 0$,
so that ${{\nu}}_h^*=\rho_h$,
consistently with the known fact that the distribution of the expected degree $h$'s is equal to $\rho_h$ 
\cite{CaldarelliHidden,BogunaHidden,NewmanHidden,SatorrasHidden}. 

However, if we consider more general cases as, like \textit{e.g.}, RGG, $p(h_i,h_j)$ is a ``rigid'' or ``soft'' 
function of the relative distances between nodes $i$ and $j$,
and we are not able to directly control $\avg{k}$ from the parameters of $p(h,h')$. 
More precisely, we cannot reach the desired $\avg{k}$ by only tuning the parameters in $p(h,h')$ 
(for example by using the sole Eq. (\ref{ptilde})).
We are then forced to use
a finite chemical potential $\mu$. If $\mu\neq 0$, we have $\sigma_{h,h'}\neq 0$ and then ${{\nu}}_h\neq \rho_h$; \textit{i.e.},
the system is not in equilibrium with respect to $\rho_h$.
To evaluate $\mu$,
from Eqs. (\ref{L}), (\ref{SQ}) and (\ref{sp}), we have
\begin{eqnarray*}
\label{LQ}
&& \mediaT{L}=-N\frac{\partial f(\{{{\nu}}_h^*\})}{\partial \mu}, \quad
\mathrm{where} \quad \nonumber \\
&&
-\frac{\partial f}{\partial \mu}=\sum_h\frac{-\partial f}{\partial {{\nu}}_{h}}
\frac{\partial {{\nu}}_{h}}{\partial \mu}+\frac{1}{2}N\sum_{h,h'}{{\nu}}_h{{\nu}}_{h'}\frac{\partial\sigma_{h,h'}}{\partial \mu},
\end{eqnarray*}    
which, when calculated at the saddle point ${{\nu}}_h={{\nu}}_h^*$, simplifies as
\begin{eqnarray*}
\label{LQ1}
&&-\frac{\partial f(\{{{\nu}}_h^*\})}{\partial \mu}=\frac{1}{2}N\sum_{h,h'}{{\nu}}_h^*{{\nu}}_{h'}^*\frac{\partial\sigma_{h,h'}}{\partial \mu}, \quad
\mathrm{with} \quad \nonumber \\
&&
\label{LQ3}
\frac{\partial \sigma_{h,h'}}{\partial \mu}=p_{h,h'}e^{\mu-\sigma_{h,h'}}.
\end{eqnarray*}    

In conclusion, we have to solve the systems of Eqs.
\begin{subequations}
\label{Global}
\begin{flalign}
\label{spq1}
{{\nu}}_h=\frac{\rho_he^{N\sum_{h'}{{\nu}}_{h'}\sigma_{h,h'}}}{\sum_{h''}\rho_{h''}e^{N\sum_{h'}{{\nu}}_{h'}\sigma_{h'',h'}}}, \\
\label{LLD}
\mediaT{L}=\frac{1}{2}N^2\sum_{h,h'}{{\nu}}_h{{\nu}}_{h'}p_{h,h'}e^{\mu-\sigma_{h,h'}}.
\end{flalign}
\end{subequations}   
or, when $\mathcal{D}$ is continuum
\begin{subequations}
\label{GlobalC}
\begin{flalign}
\label{spq1C}
{{\nu}}(h)=\frac{\rho(h)e^{N\int dh' {{\nu}}(h')\sigma(h,h')}}{\int dh''\rho(h'')e^{N\int dh' {{\nu}}(h')\sigma(h',h'')}}, \\
\label{LL}
\mediaT{L}=\frac{1}{2}N^2\int dh dh' {{\nu}}(h){{\nu}}(h')p(h,h')e^{\mu-\sigma(h,h')}.
\end{flalign}
\end{subequations}   
The sums in (\ref{Global}) and the integrals in (\ref{GlobalC}) are understood to run over the set $\mathcal{D}$.
In the continuum, the system (\ref{GlobalC}) is a functional system. In such a case the discrete system (\ref{Global})
provides a numerical approximation to (\ref{GlobalC}) whose accuracy grows with the number of points $N_s$ used to discretize $\mathcal{D}$.
The system (\ref{Global}) consists of $|\mathcal{D}|+1$ Eqs. in the $|\mathcal{D}|+1$ unknowns $\left(\{{{\nu}}_h\};\mu\right)$, with
the $|\mathcal{D}|+1$ inputs $\left(\{\rho_h\};\mediaT{L}\right)$.
In the following, with some abuse of notation, sometimes we will write $|\mathcal{D}|=N_s$. It will be clear from the context whether we are referring to 
a model with discrete value hidden variables, or to the discrete approximation of a model in the continuum.  

It is useful to observe that Eq. (\ref{LL}) (and similarly Eq. (\ref{LLD})) can also be written as
\begin{eqnarray}
\label{LL1} 
\mediaT{L}=\frac{1}{2}N^2\int dh dh' {{\nu}}(h){{\nu}}(h'){{q}}_\mu(h,h'),
\end{eqnarray}
where we have introduced the probability
\begin{eqnarray}
\label{ptilde} 
&&{{q}}_\mu(h,h')=p(h,h')e^{\mu-\sigma(h,h')}\nonumber \\
&&=\frac{p(h,h')e^{\mu}}{p(h,h')e^{\mu}+1-p(h,h')}.
\end{eqnarray}
Eqs. (\ref{Global})-(\ref{ptilde}) make clear that we are dealing with an ensemble duality: 
\begin{eqnarray}
\label{Dual} 
Ensemble\left\{\bm{p},\bm{\rho},\mu\right\}=Ensemble\left\{{\bm{q}}_\mu,{\bm{\nu}}_\mu,0\right\}
\end{eqnarray}
The duality (\ref{Dual}) says that, given $N$ and $\mathcal{D}$, in order to draw a network with
a desired average connectivity $\avg{k}=~2\avg{L}/N$, 
we need to look for a proper distribution of nodes ${\bm{\nu}}_\mu(h)$ and for a proper distortion of $p(h,h')$ 
via ${{q}}_\mu(h,h')$
to favor ($\mu>0$) or unfavor ($\mu<0$) the probability to have a link between $h$ and $h'$.
Since (\ref{Dual}) has been obtained by using a saddle point technique, it is exact only in the thermodynamic 
limit $N\to \infty$, and
the $Ensemble\left\{{\bm{q}}_\mu,{\bm{\nu}}_\mu,0\right\}$ contains only the typical graphs of 
the $Ensemble\left\{\bm{p},\bm{\rho},\mu\right\}$.

\section{A toy example}
Let us consider the classical random graph.
Within the hidden-variables formalism one has
\begin{eqnarray}
\label{CT}
&& p(h,h')= \frac{c}{N},
\end{eqnarray}
\begin{eqnarray}
\label{rhoT}
\rho(h)=1, \quad h\in \mathcal{D},
\end{eqnarray}
where $\mathcal{D}=[0,1]$ is an interval and $c>0$ is a constant.
Due to the fact that $p(h,h')$=constant, we have also $\sigma(h,h')$=constant 
and Eqs. (\ref{spq1C})-(\ref{LL}) simply give
\begin{subequations}   
\begin{flalign}
\label{rhoT1}
{{\nu}}_\mu(h)=\rho(h), \\
\label{rhoT2}
\avg{k}=\frac{ce^\mu}{1+\frac{c}{N}(e^\mu-1)}\simeq ce^\mu.
\end{flalign}
\end{subequations}   
Eqs. (\ref{rhoT1})-(\ref{rhoT2}) tell us that we can reach any desired average connectivity $\avg{k}$ without
changing $\rho(h)$, and that $\mu\neq 0$ simply amounts to a renormalization of the constant $c$.
Therefore,  in this example the duality relation (\ref{Dual}) does not add any new information.  

\section{Application to Random Geometric Graphs}
\subsection{General definition of RGG}
In RGG the $h$'s represent the positions of the nodes and
two nodes are connected or not according to 
\begin{eqnarray}
\label{rule0}
p_{h,h'}=\left\{
\begin{array}{l}
1, \quad \mathrm{if~}h\mathrm{~and~}h'\mathrm{~satisfy~a~geometric~rule}, \\
0, \quad \mathrm{otherwise}.
\end{array}
\right.
\end{eqnarray}   
Eqs. (\ref{rule0}) plugged into Eqs. (\ref{SQ}) give 
\begin{eqnarray}
\label{rule01}
\sigma_{h,h'}=\left\{
\begin{array}{l}
\mu, \quad \mathrm{if~}h\mathrm{~and~}h'\mathrm{~satisfy~a~geometric~rule}, \\
0, \quad \mathrm{otherwise}.
\end{array}
\right.
\end{eqnarray}   
Eq. (\ref{rule0}) represents rigid RGG. More in general, we might consider
soft RGG where Eq. (\ref{rule0}) is replaced by
\begin{eqnarray}
\label{rules}
p_{h,h'}=\left\{
\begin{array}{l}
a, \quad \mathrm{if~}h\mathrm{~and~}h'\mathrm{~satisfy~a~geometric~rule}, \\
1-a, \quad \mathrm{otherwise},
\end{array}
\right.
\end{eqnarray}   
where $0<a<1$ can be a constant or any function of the node positions $h$ and $h'$.

Particularly important are the RGG defined over a continuous domain $\mathcal{D}$ equipped
with a distance. 
In the rigid case ($a=1$ in Eq. \ref{rules}), for any sprinkle of nodes, 
there is one single graph obtained by connecting
all pairs of nodes whose distance is at most $r$, $r>0$ being a fixed parameter, called threshold. 
Nodes are sprinkled according to the given PDF $\rho(h)$, and the support of $\rho(h)$ defines the domain $\mathcal{D}$.
A natural choice for $\rho(h)$ is the uniform PDF $\rho(h)$=constant, and this will be the case in the following
numerical examples. However, for practical and theoretical reasons
it is convenient to think of $\rho(h)$ as an arbitrary PDF. 

We stress that, for rigid RGG  we have $q_\mu(h,h')=p(h,h')$ but $\nu_\mu(h)\neq \rho(h)$.

\subsection{Solving the Saddle Point Eqs; Dynamics of RGG}
Suppose for simplicity that $\mu$ is known. Given any $\rho_h$, which plays the role of an initial density ${{\nu}}_h(t=0)=\rho_h$, Eqs. (\ref{spq1})
leads to a natural dynamics over the ${{\nu}}_h$'s. For example, we can define a discrete-time dynamics by simple iteration of the ${{\nu}}_h$'s: 
\begin{eqnarray}
\label{spqd}
&& {{\nu}}_h(t+1)=\frac{\rho_he^{N\sum_{h'}{{\nu}}_{h'}(t)\sigma_{h,h'}}}{\sum_{h''}\rho_{h''}e^{N\sum_{h'}{{\nu}}_{h'}(t)\sigma_{h'',h'}}}, \quad \nonumber \\
&& {{\nu}}_h(t=0)=\rho_h.
\end{eqnarray}
Under the hypothesis that there exists a stable fixed point, the above iteration offers a way to solve Eqs. (\ref{spq1}) iteratively, 
There are many ways to define the dynamics, but all make use of Eqs. (\ref{spq1}) and therefore have the same stationary solutions.
In general, if $\mu\neq 0$, Eqs. (\ref{spq1}) can generate a ``diffusion'' with asymptotic solution toward ${{\nu}}_h(t=\infty)={{\nu}}_h$. 
This picture is very appealing for RGG. However, we have to keep in mind that we need to work with a large value of $|\mathcal{D}|=N_s$ in order
to recover a diffusion-like trend. 
In fact, if we choose for $\rho_h$ a point-like distribution:
\begin{eqnarray}
\label{point}
\rho_h=\delta_{h,h_0},
\end{eqnarray}   
it is immediate to check that Eqs. (\ref{spq1}) are solved with
\begin{eqnarray}
{{\nu}}_h=\delta_{h,h_0}.
\end{eqnarray}   
More in general, ${{\nu}}_h$ is always of the form 
\begin{eqnarray}
{{\nu}}_h=\chi_{\mathcal{D}}(h)f_h,
\end{eqnarray}   
where $\chi_{\mathcal{D}}(h)$ is the characteristic function of the set $\mathcal{D}$ and $f_h$ a suitable vector.
It is then clear that for small values of $|\mathcal{D}|$, we cannot have a diffusion like trend.
We expect to see a diffusion when $|\mathcal{D}|\sim N^\alpha$, with $\alpha>0$.
RGGs, however, are usually defined over a continuous domain $\mathcal{D}$ and for an exact solution we should solve the functional system (\ref{GlobalC}).
As anticipated, a numerical solution relies on a discretization of $\mathcal{D}$ 
by using a large enough number of points $N_s$.
Given $N$, the numerical solution will approach the exact solution in the limit $N_s/N \to \infty$. 
However, for all practical aims a finite ratio
provides excellent approximations. 
Note that solving (\ref{Global}) by using the iteration (\ref{spqd}) produces only the solutions
that are minima (local or global) of the free energy density  (\ref{SQ1}).

\comment{Finally, we point out that, although there might exist smarter methods to solve (\ref{Global}), traditional population dynamics here cannot be used.
In fact, in population dynamics 
the iteration does not run for any $h$ and the sums in (\ref{spqd}) are 
sampled for randomly selected sites and their respective neighbors (here two sites are neighbors if $\sigma_{h,h'}\neq 0$). 
Population dynamics returns the exact iteration when $N_s\to \infty$.
However, it should be clear that such a sampling relies on the statistical knowledge of the first neighbors of each site that must be finite.
In our case we do not have access to this knowledge when $N_s\to \infty$. In fact, if $\mathcal{D}$ is continuum, 
given $h$, the number of site-neighbors of $h$ diverges for $N_s\to \infty$. 
In conclusion, we are forced to rely to the full exact iteration (\ref{spqd}) with finite $N_s$.}

\subsection{RGG on the circle}
We can deepen our understanding with the simplest RGG: the circle. In this case $\mathcal{D}=[0,2\pi)$. Given the number of nodes $N$,
we choose a discretization of $\mathcal{D}$ with $N_s$ equidistant sites; in other words the $N$ nodes can occupy any of the $N_s$ sites
whose positions are given by their 
angles $\theta_h$, $h=1,\ldots,N_s$. The $N$ node are randomly sprinkled through a given distribution $\rho_h$.
Then we connect two points according to the following probability
\begin{eqnarray}
\label{rule}
p(\theta_h,\theta_{h'})=\left\{
\begin{array}{l}
a, \quad ||\theta_h-\theta_{h'}||<{r}, \\
1-a, \quad ||\theta_h-\theta_{h'}||\geq {r},
\end{array}
\right.
\end{eqnarray}   
where $a$ and ${r}$ are constant, and $||\theta-\theta'||=\pi-|\pi-|\theta-\theta'||$ is the distance on the circle.
We set
\begin{eqnarray}
\label{d0}
{r}=\frac{4\pi}{N}.
\end{eqnarray}   
Eqs. (\ref{rule}) plugged into Eqs. (\ref{SQ}) give 
\begin{eqnarray}
\label{rule1}
\sigma(\theta_h,\theta_{h'})=\left\{
\begin{array}{l}
\ln\left[ae^\mu+1-a\right], \quad ||\theta_h-\theta_{h'}||<{r}, \\
\ln\left[(1-a)e^\mu+a\right], \quad ||\theta_h-\theta_{h'}||\geq {r}.
\end{array}
\right.
\end{eqnarray}   
If for $\rho_h$ we choose the uniform distribution 
\begin{eqnarray}
&&\rho_h=\rho(\theta_h)=\frac{1}{N_s}, \quad \nonumber \\
&&\theta_h\in \left\{0,\frac{2\pi}{N_s},\frac{4\pi}{N_s},\ldots,\frac{(N_s-1)2\pi}{N_s}\right\},
\end{eqnarray}   
it is immediate to check that ${{\nu}}_h=\rho(\theta_h)$, independently from $\mu$ and $a$
(this holds for any homogeneous RGG: if $\rho_h$ is the uniform distribution, ${{\nu}}_h=\rho_h$).
However, for $\rho(\theta_h)$ we can choose a distribution which differs slightly from the uniform one:
\begin{eqnarray}
\label{Random}
\rho_h =A(1+\epsilon_h)
\end{eqnarray}   
where $\epsilon_h$ are uniformly distributed random variables with mean 0 and finite variance, and $A$ is a normalization constant.
In general, if we use Eq. (\ref{spqd}) with the initial distribution (\ref{Random}),  
the system may evolve toward a distribution which is different from the uniform one.
In principle, any initial distribution (provided not equal to the uniform one) can be used to solve the system (\ref{Global}).
In fact, we find that via (\ref{spqd}) they all tend to the same solution for (\ref{Global}), provided
they have the same support in $\mathcal{D}$.

\begin{figure}[htbp]
{\includegraphics[height=1.6in]{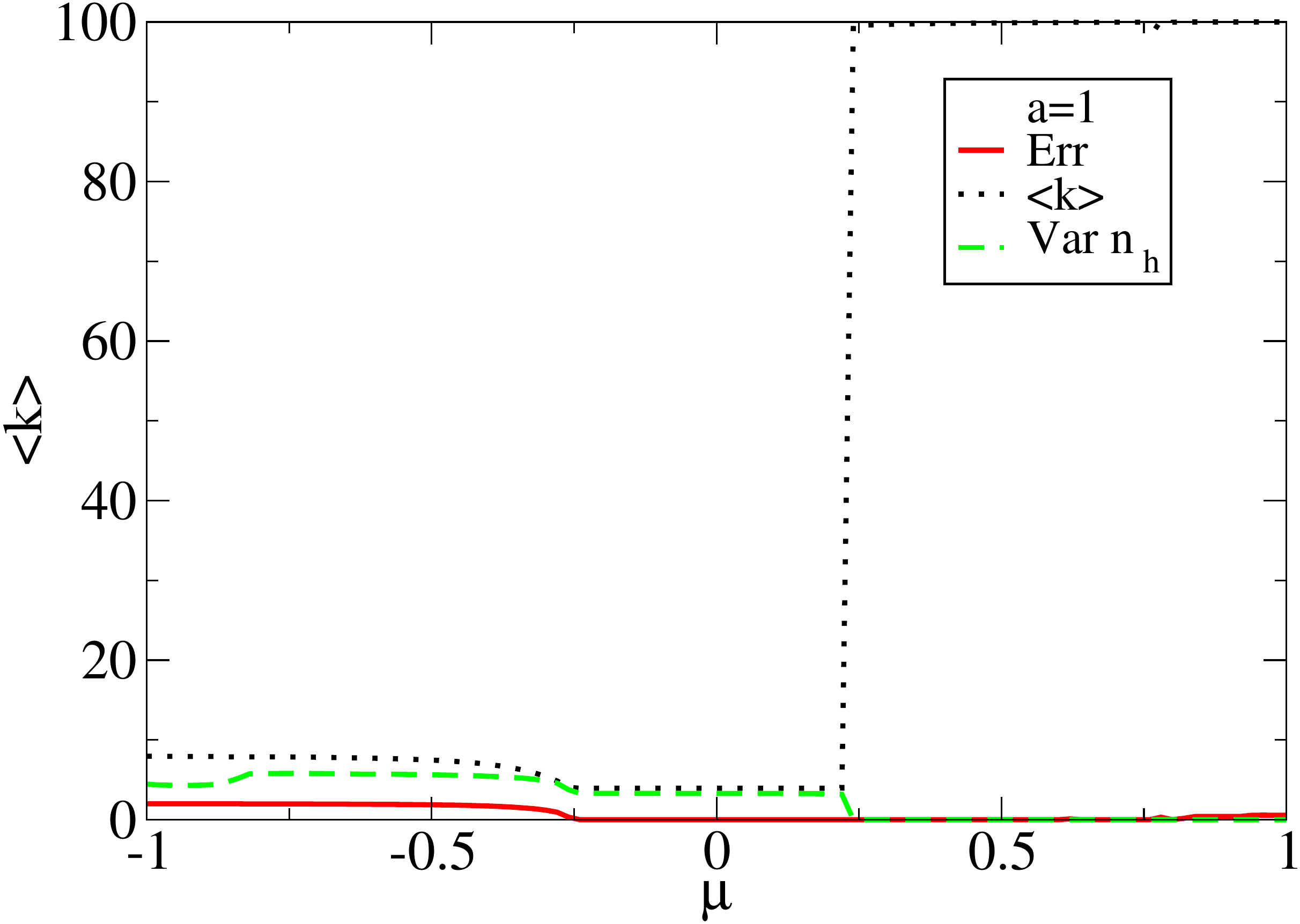}}\\
{\includegraphics[height=1.6in]{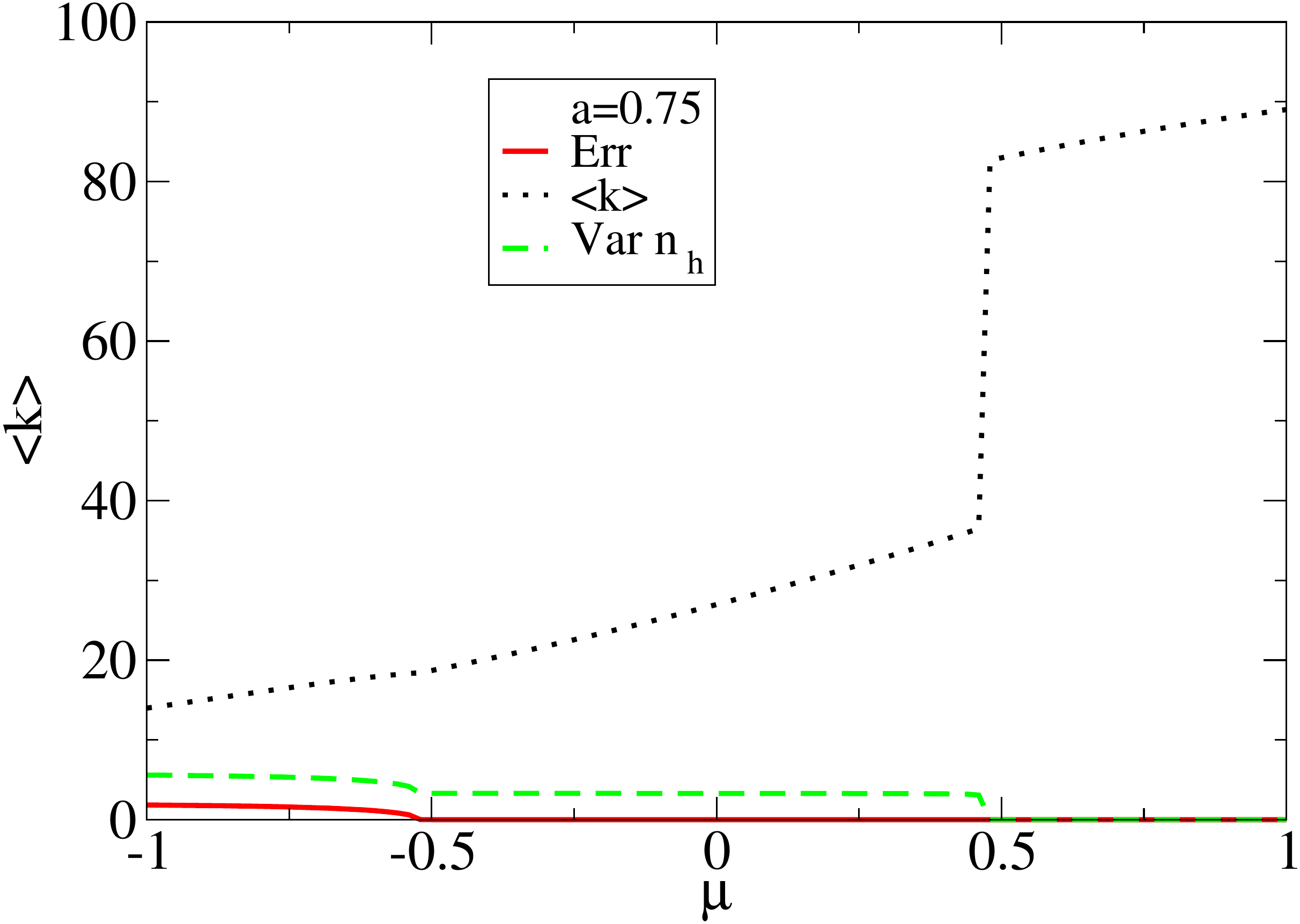}}\\
{\includegraphics[height=1.6in]{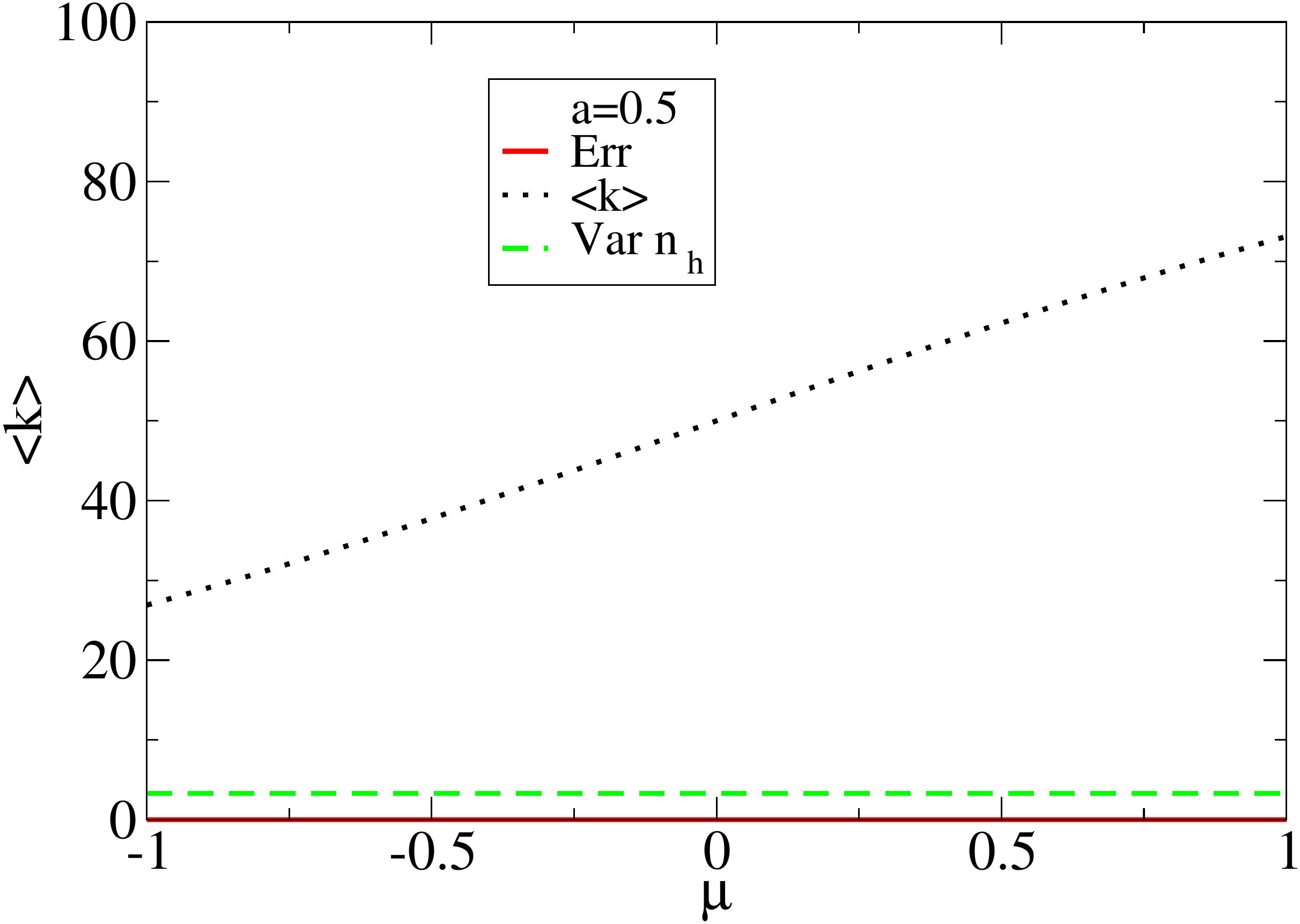}}\\
{\includegraphics[height=1.6in]{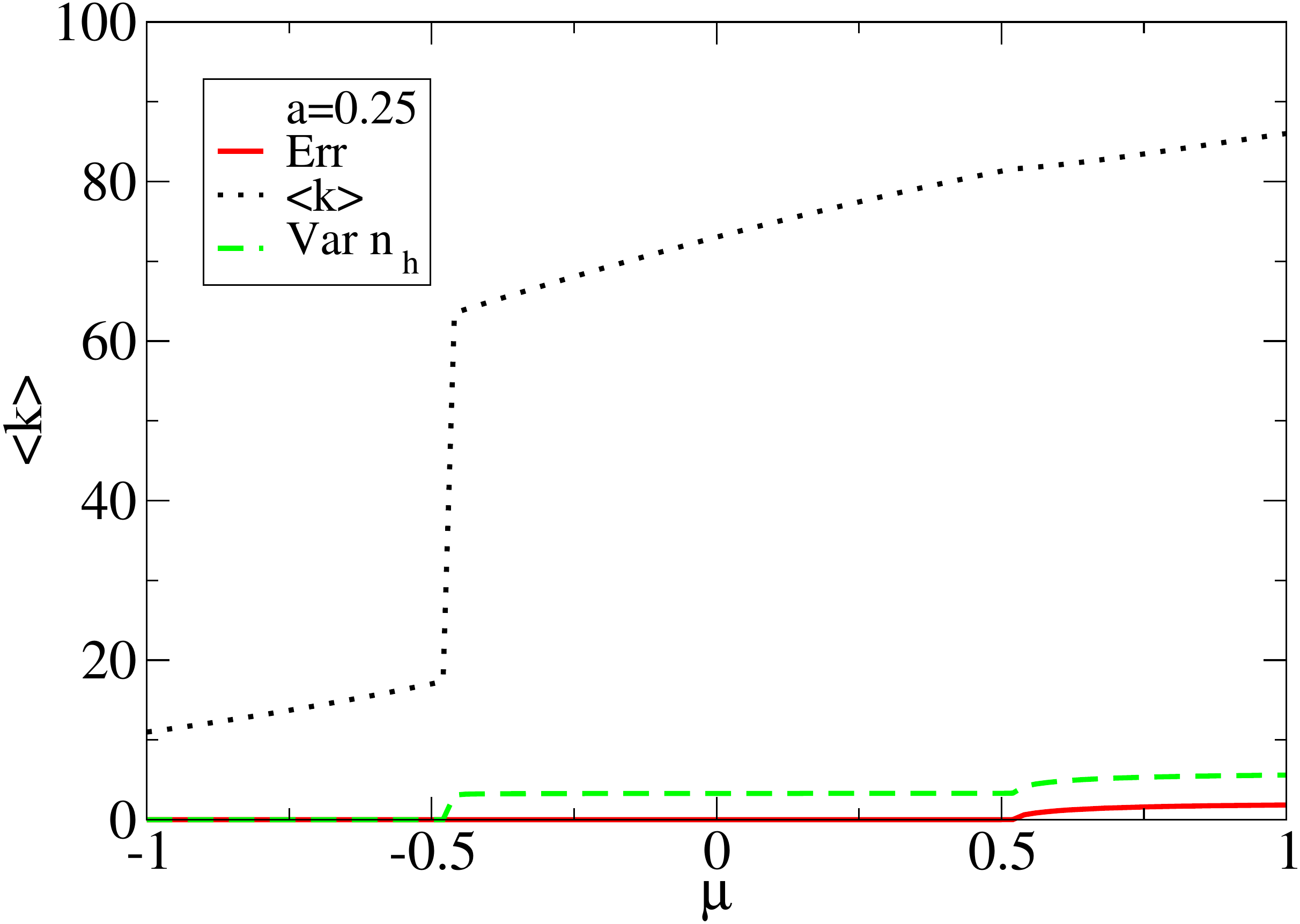}}\\
{\includegraphics[height=1.6in]{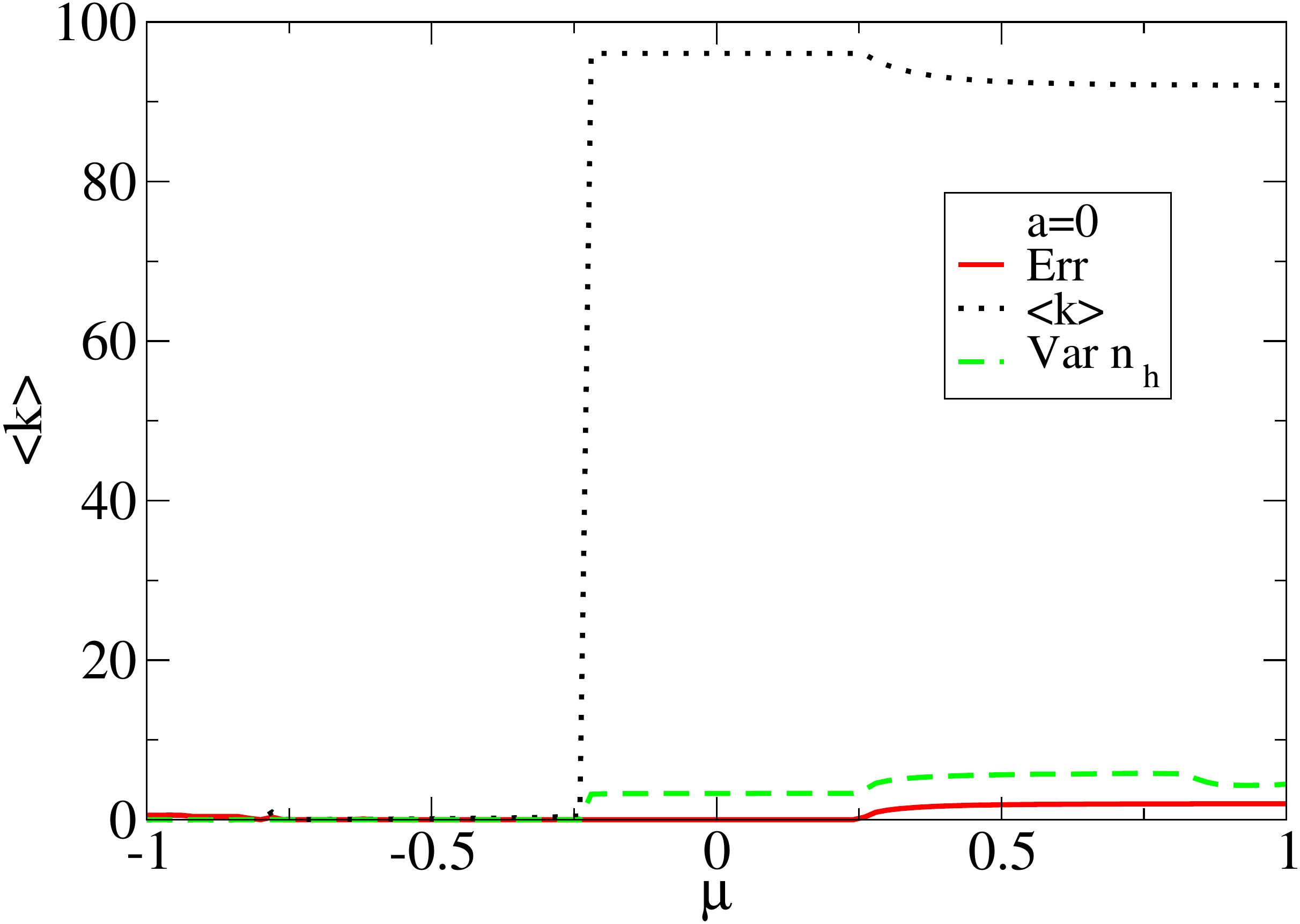}}\\
\caption{(Color online) Analysis of the RGG (\ref{rule})-(\ref{d0}) for several values of $a$ (here $\nu_h$ stands for $\nu_\theta$). 
We plot $\avg{k}$ 
\textit{vs} the chemical potential $\mu$. 
Err=$\sum_h |{{\nu}}_h(t+1)-{{\nu}}_h(t)|$ provides an evaluation of the existence of solutions: For Err$>0$ there are no solutions; Var$~{{\nu}}_h$ is the variance of ${{\nu}}_h(t)$.
Data obtained by solving the discrete system (\ref{Global}) 
via the iteration (\ref{spqd}) with $t=200$ iterations. 
Here the number of nodes is $N=100$,
whereas the discretization parameter is set to $N_s=1000$. 
The RGG is exactly reproduced only in the limit $N_s\to \infty$, however, for our numerical aims
$N_s=1000$ is large enough and, on this scale, larger values of $N_s$ do not produce serious differences.
\label{fig2}
}
\end{figure}

\begin{figure}[htbp]
{\includegraphics[height=1.8in]{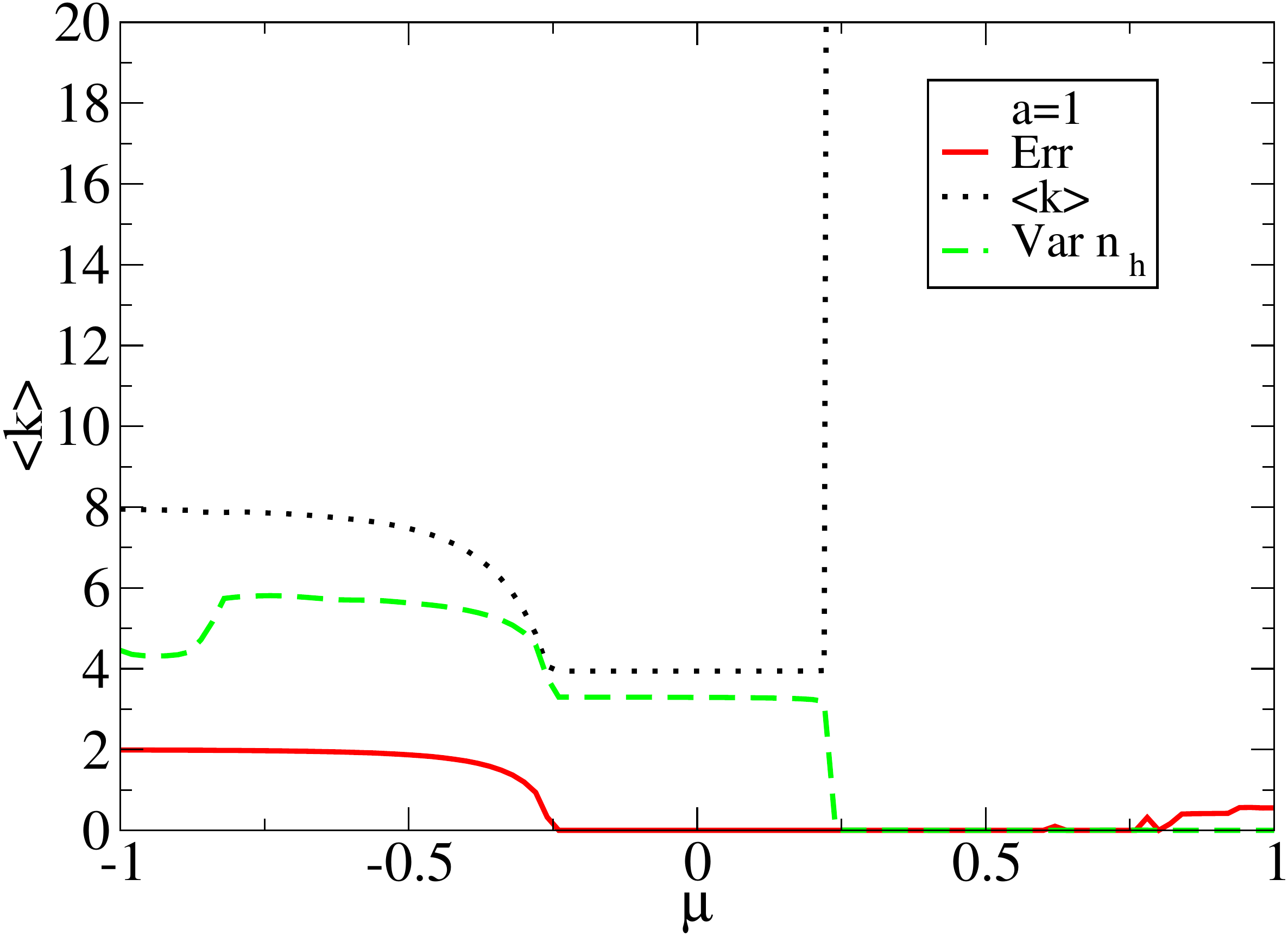}}\\
{\includegraphics[height=1.8in]{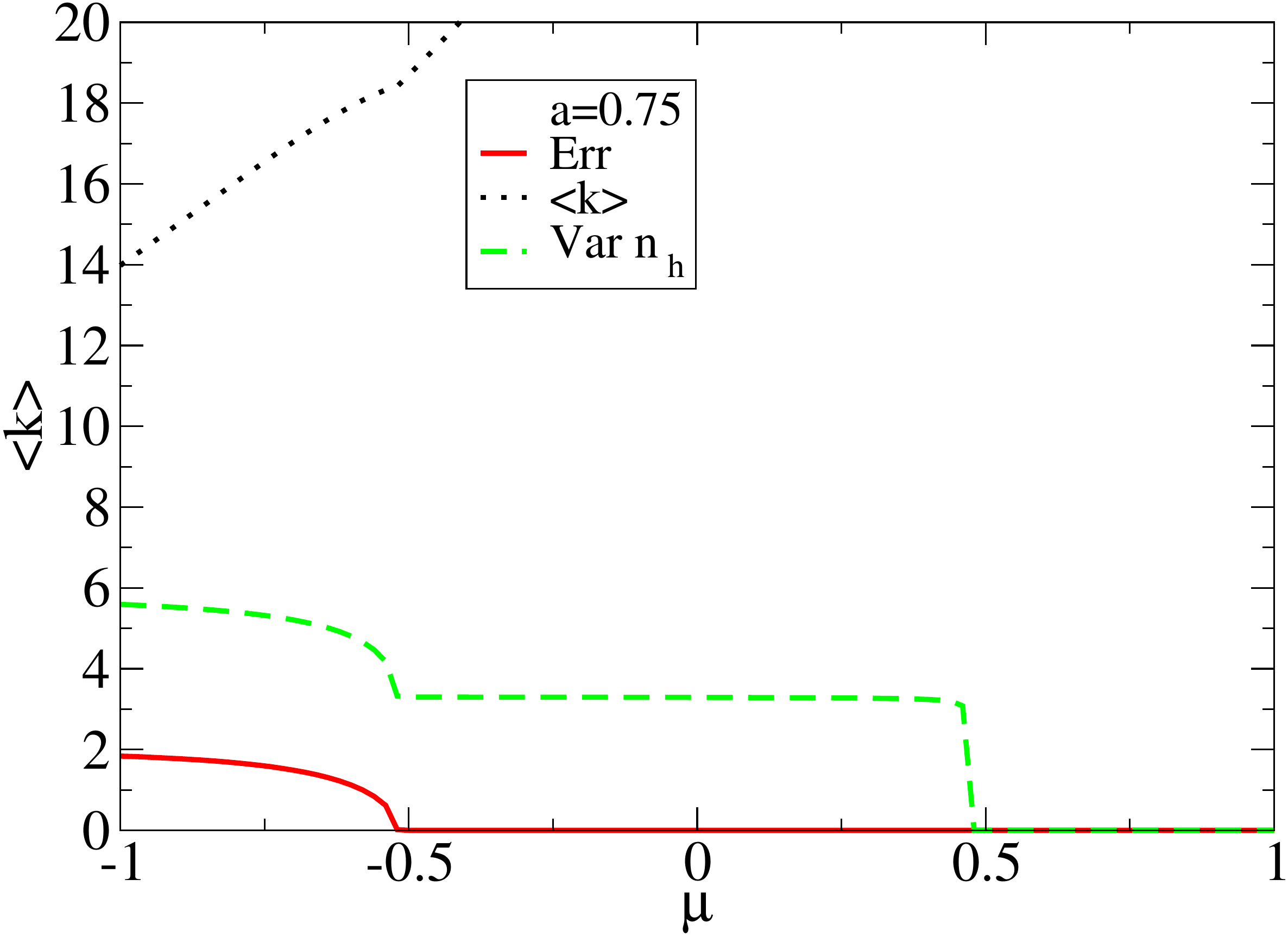}}\\
{\includegraphics[height=1.8in]{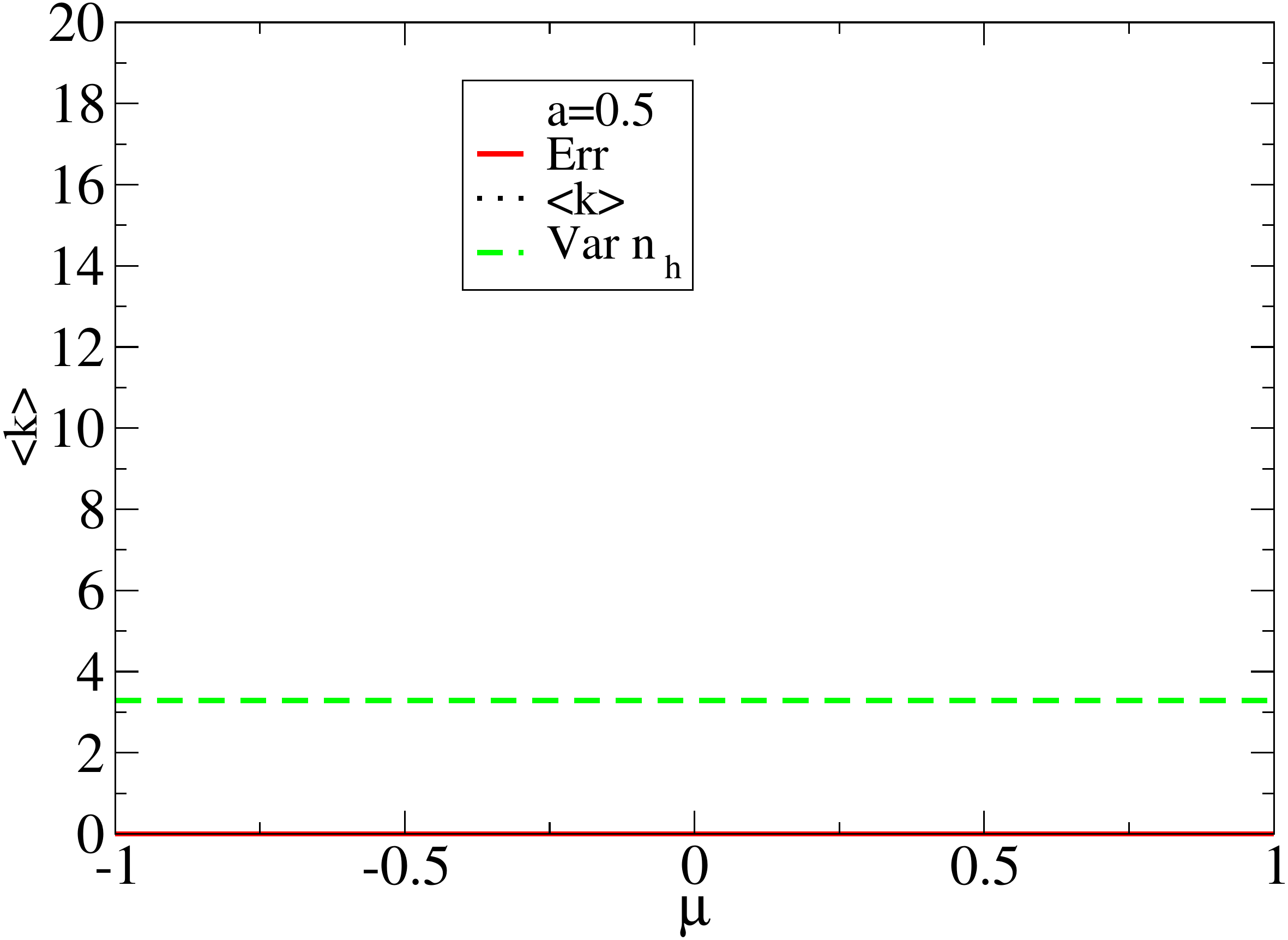}}\\
{\includegraphics[height=1.8in]{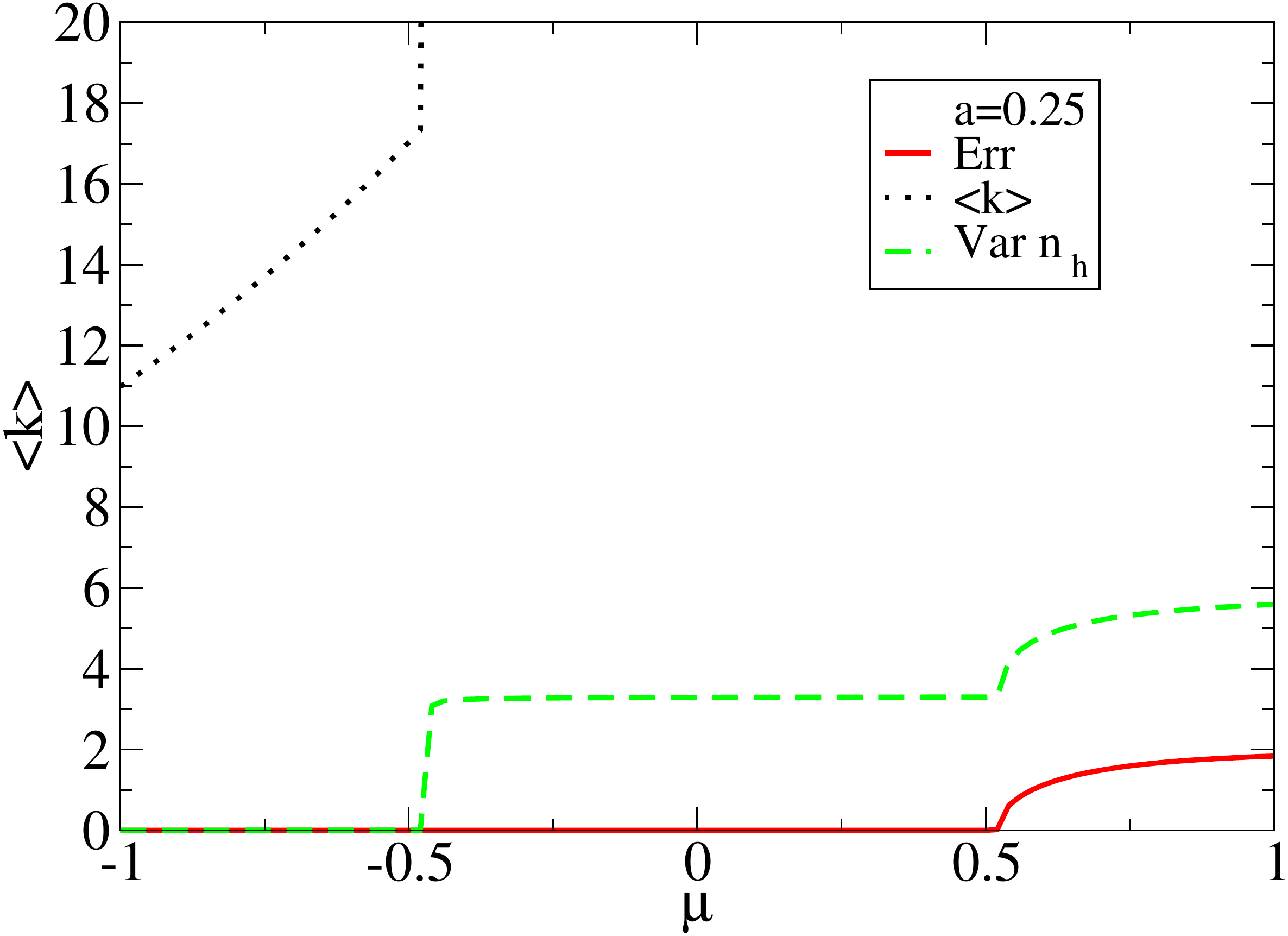}}\\
{\includegraphics[height=1.8in]{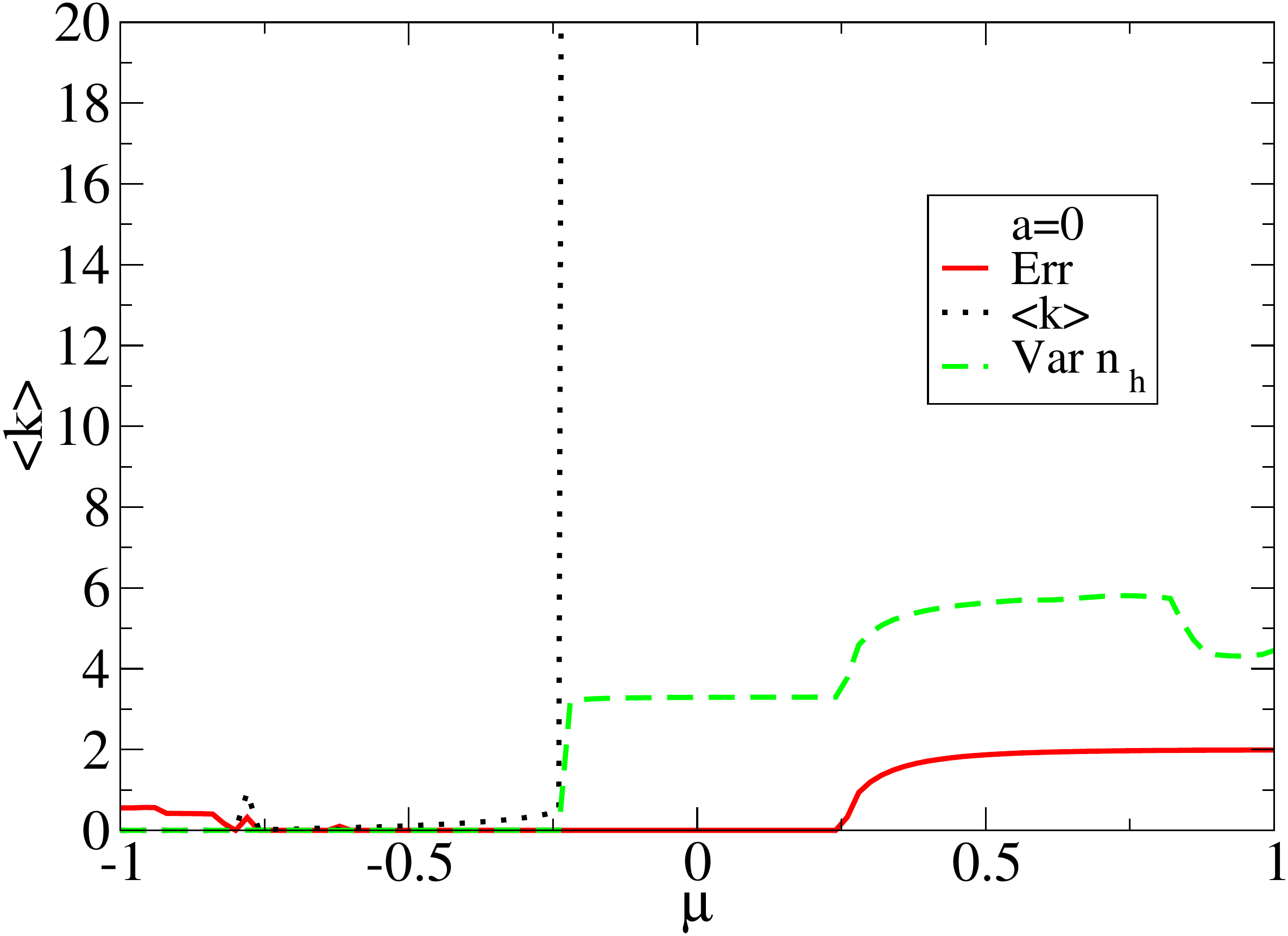}}
\caption{(Color online) Analysis of the RGG (\ref{rule})-(\ref{d0}) for several values of $a$ (here $n_h$ stands for $\nu_\theta$). 
Enlargements of the plots of Figs. \ref{fig2}.
\label{fig2b}
}
\end{figure}

\begin{figure}[htbp]
{\includegraphics[height=2.2in]{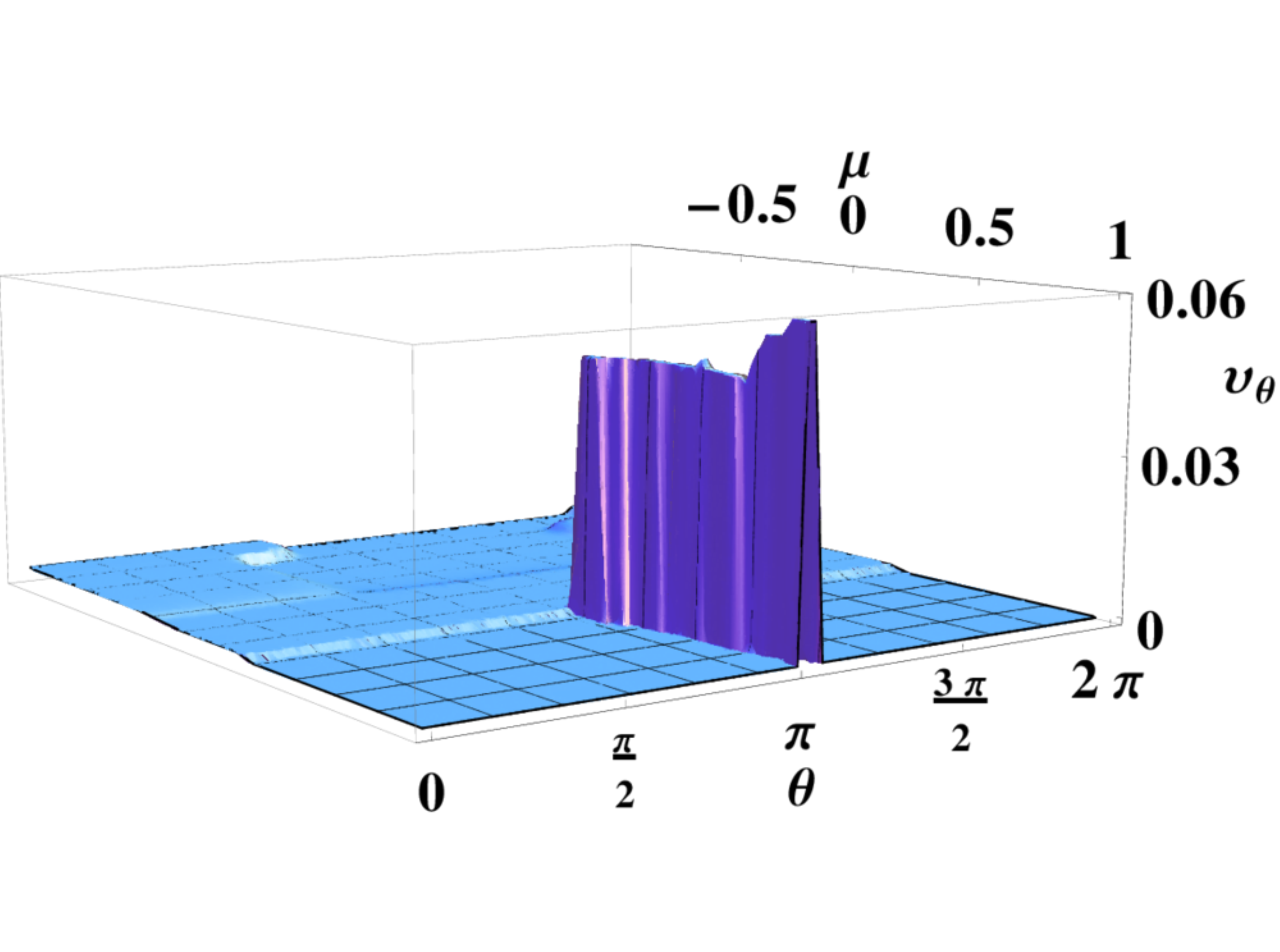}}
\caption{(Color online) Distribution ${{\nu}}_\theta$ for the RGG (\ref{rule})-(\ref{d0}) with $a=1$, 
as a function of the angle $\theta$ for each value of the chemical potential $\mu$. 
Plot obtained with $N=100$, $N_s=1000$ and t=$t=200$, as for Fig. \ref{fig2}.
Note that, in the localized regime, there is an high degeneracy of solutions, 
since the $N$ points can get localized in any small region of the interval $[0,2\pi)$. 
Up to this degeneracy, different initial conditions (except the trivial one (\ref{point})) lead to the same solution.
Finite size effects can be seen in Figs. \ref{fig0}.
\label{fig1}}
\end{figure}

We expect that, the larger is $\avg{k}=\avg{2L}/N$, the larger is $\mu$.
However, it is a priori difficult to know whether, for any $\avg{k}$, there corresponds a solution
for (\ref{Global}). In fact, in general this is not the case, furthermore, we find out  
that $\avg{k}$, as a function of $\mu$, undergoes a $\mathop{O}(N)$ jump 
in correspondence of a critical $\mu_c$, as depicted in Figs. \ref{fig2}-\ref{fig1}
where we analyze several values of $a$.

\subsection{The general scenario: a geometry-induced condensation of nodes and links}
\begin{figure}[htbp]
{\includegraphics[height=2.0in]{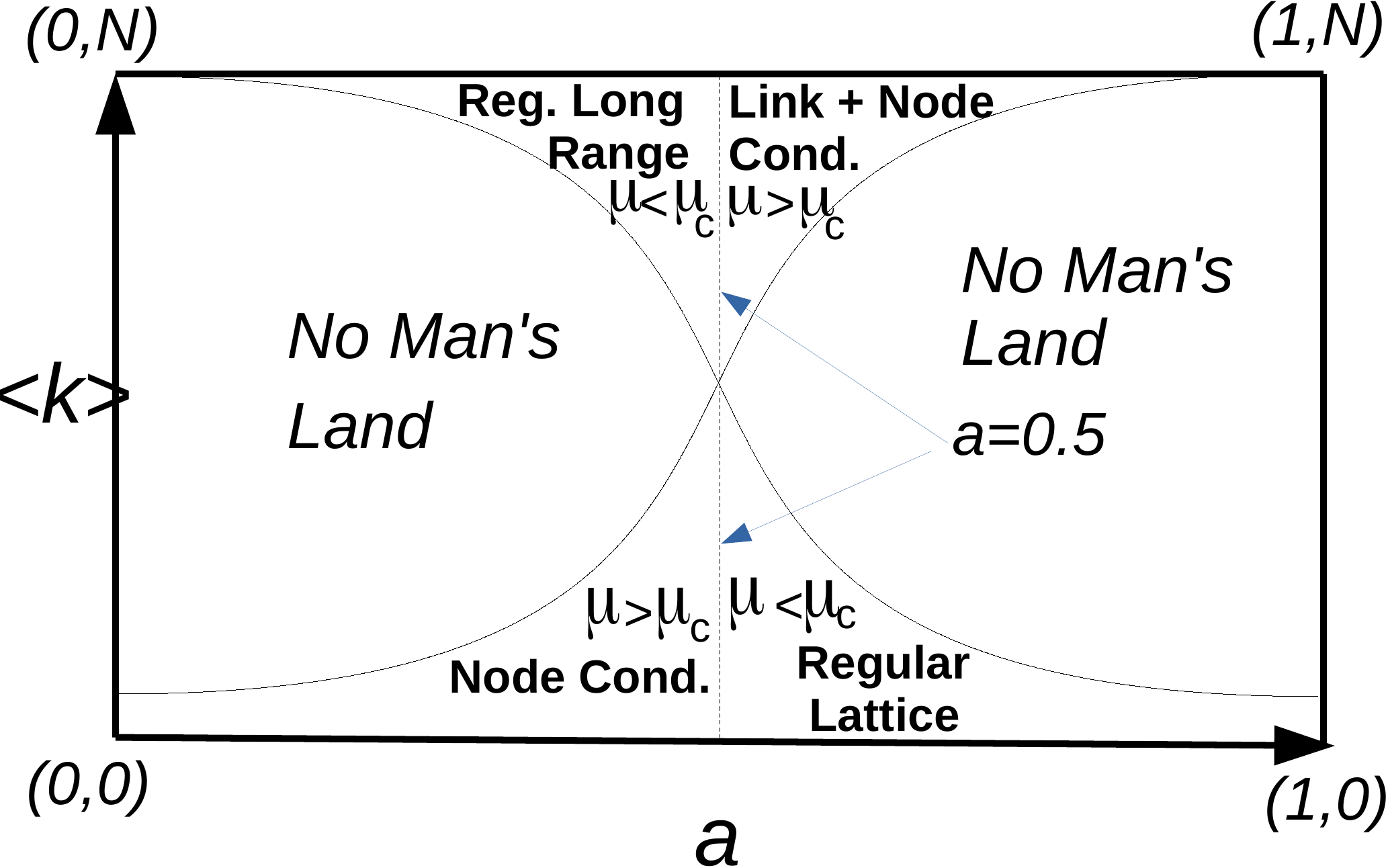}}
\caption{\comment{(Color online) Pictorial phase diagram of a Euclidean RGG in the plane $(a,\avg{k})$.
Given a threshold $r$, two nodes are connected with probability $a$ if their distance is at most $r$. 
The regions labeled ``No Man' s Lands'' are regions populated by only atypical (\textit{i.e.}, exponentially rare, or hard solutions) graphs. 
There are four stable phases (or easy solutions): a condensate and a regular region, for both $a>1/2$ and $a<1/2$. 
Case $a>1/2$ (close nodes are connected with probability $a>1/2$): 
the condensate zone is the
region where typical graphs are characterized by a condensation of nodes and links, with the resulting graphs nearly fully connected, whereas
the regular zone is the region where typical graphs have nodes nearly uniformly distributed and with a minimal connectivity, as in a regular lattice.
Condensate and regular regions are characterized by $\mu>\mu_c$ and $\mu<\mu_c$, respectively.
For $a<1/2$ (distant nodes are connected with probability $1-a>1/2$) the picture is reversed, with condensed nodes having minimal connectivity, and uniformly distributed nodes having maximal
connectivity (due to long range connections), for $\mu>\mu_c$ and $\mu<\mu_c$, respectively.
At $a=0.5$ there is no geometry and no phase transition. 
Notice that $\mu_c$ depends on $a$. We point out also that, in the exact phase diagram, the minimal and maximal average connectivities
of the cases $a>0.5$ are slightly smaller than the minimal and maximal average connectivities of the cases $a<0.5$. 
The reason for this difference is due to a simple geometrical argument whose details are left to the reader.
}   
\label{fig3}
}
\end{figure}

The phase transition depicted in Figs. \ref{fig0}-\ref{fig1} 
is perfectly compatible with the geometric interpretation and
we can summarize the general scenario as follows. Given $N$ nodes and a rigid ($a=1$) RGG defined over a continuous domain $\mathcal{D}$, 
we are free to arrange the $N$ nodes in any way, with two opposite limit regimes: If nodes are uniformly distributed over 
$\mathcal{D}$, $\avg{k}=\mathop{O}(1)$, while if nodes are localized in a small subset of $\mathcal{D}$, $\avg{k}=\mathop{O}(N)$.
However, our analysis shows that, when $N\to\infty$, the probability to find a configuration out of these two regimes tends to 0, and the two regimes
become two phases separated by a first order phase transition: a uniformly diluted/regular phase, and a condensate of nodes and links in the other phase.
Such a transition is triggered by the geometry whose strength can be tuned by the value of the parameter $0<a\leq 1$. 
For $a>1/2$ the two regimes are separated by the geometry: Close nodes have an high number of links.
For $a<1/2$ the two regimes are separated by an anti-geometry: Close nodes have a low number of links. 
Finally, for $a=1/2$
there is no geometry and no phase transition. \comment{See Fig. \ref{fig3} for a qualitative phase diagram description.}
We stress that this phase transition scenario has nothing to share with the percolation phenomena
(in Appendix \ref{appB} we show how to deal with the percolation problem within our framework). 
In fact, as we shall prove later by a combinatorial argument,
the scenario does not depend on the details of the model, nor on the dimension $d$, or on the particular value $r$ chosen, which can be changed
to scale in the very sparse or dense regimes 
(as we have numerically checked by replacing $r$ in Eq. (\ref{d0}) with $r \propto 1/N^{2}$ and $r \propto \ln(N)/N$, respectively) 
without affecting the qualitative features of the phase transition.

\subsection{Combinatorial argument}
From a graph construction viewpoint, one might wonder why there is this jump in $\avg{k}$. 
In other words, if we are free to lay down nodes
as we like, why are only certain values of $\avg{k}$ visible when $N\to\infty$ ?
Given a RGG with domain $\mathcal{D}$, let us consider for simplicity the rigid case $a=1$. 
In this case, given the connectivity $\avg{k}$ (or, equivalently, given $\mu$), the node-positions are all equiprobable.
It is also convenient to proceed with a discretized version of $\mathcal{D}$, with $|\mathcal{D}|=N_s=\alpha N$, where $\alpha \gg 1$ is an arbitrarily large constant. 
When we look at all the possible realizations of the graphs, we start with the graph with minimal connectivity $\avg{k}=\mathop{O}(1)$, 
where the nodes are uniformly distributed and this configuration is roughly unique (0 degeneracy). 
Next, we consider the opposite case where the graph has maximal connectivity $\avg{k}=N$, where the nodes are densely localized 
over a small region (average distance smaller than ${r}$), and this configuration has degeneracy which goes like $N$. 
Let us now consider the configurations in which there are two separated highly dense regions having $N/2$ nodes each. 
In this latter case we have $\avg{k}=N/2$, and the degeneracy of such possible configurations goes like $N^2$. According to this counting, we should then expect
to see this latter case, where $\avg{k}=N/2$, as highly more probable than the former case, where $\avg{k}=N$, which is the opposite of what we have
seen in the previous paragraph (see top panel of fig. \ref{fig2}).  
The apparent paradox is due to the fact that such a naive counting does not take into account the perturbations from these ideal cases. 
Suppose we perturb slightly the case with $\avg{k}=N$ in the following way: We lay down all the nodes again in a single small region, yet this region has
a slightly larger extension such that the total number of links is $L_1=x \cdot N_1$, where $N_1=\binom{N}{2}$, and where $0<x<1$, with $ x \sim 1$. 
Due to the fact that $N_s$ can be set as large as we like ($\mathcal{D}$ is continuous), this is always possible, for any $x$. 
When $x \to 1$ we recover the ideal case having only one possible type of realization:
the fully connected graph. However, when $x<1$, the number of possible graphs $\mathcal{N}_1$ that we can construct in this way goes like
\begin{eqnarray}
\label{NG1}
&&\mathcal{N}_1(x)=\alpha N\binom{N_1}{L_1}\simeq \alpha N e^{N_1s(x)}, \nonumber \\ 
&&\mathrm{with}\quad \avg{k}=\frac{2L_1}{N}=Nx
\end{eqnarray}  
where the approximation holds for $N$ large, and $s(x)\geq 0$ is a ``one-particle entropic factor'' 
\begin{eqnarray}
\label{sx}
s(x)=-\ln(x)x-(1-x)\ln(1-x).
\end{eqnarray}  
Similarly, if we perturb slightly the ideal case with $\avg{k}=N/2$, in at least one of the two highly dense and separated regions, 
by introducing $N_2=\binom{N/2}{2}$, and $L_2=x\cdot N_2$, we see that 
the total number of possible graphs $\mathcal{N}_2$ that we can construct in this way goes like
\begin{eqnarray}
\label{NG2}
&& \mathcal{N}_2(x)=\binom{\alpha N}{2}\binom{N_2}{L_2}^2\simeq \binom{\alpha N}{2} e^{2N_2s(x)}, \nonumber \\ 
&&\mathrm{with} \quad \avg{k}=\frac{2L_2}{N}=\frac{N}{2}x,
\end{eqnarray}  
where $s(x)$ is again given by Eq. (\ref{sx}). 
If we put at ratio Eqs. (\ref{NG1}) and (\ref{NG2}) we get
\begin{eqnarray}
\label{ratio}
\frac{\mathcal{N}_2(x)}{\mathcal{N}_1(x)} \simeq \frac{\alpha N}{2} e^{(2N_2-N_1)s(x)}=\frac{\alpha N}{2} e^{\frac{-N^2}{2}s(x)}.
\end{eqnarray}  
We can iterate the above argument for other values of $\avg{k}$.
In particular, we can consider the configurations
having $2^q$ small equally populated regions, each having an average connectivity $\avg{k}=N/2^q$, where $q$ is an integer.
In this case the leading term of
the ratio between $\mathcal{N}_1(x)$ and $\mathcal{N}_{2^q}(x)$ goes like
\begin{eqnarray}
\label{ratioq}
\frac{\mathcal{N}_{2^q}(x)}{\mathcal{N}_1(x)} \sim \binom{\alpha N}{2^{q-1}} e^{-\frac{N^2}{2^q}s(x)},
\end{eqnarray}  
It is then clear that, as far as $q$ is finite, the configurations with $\avg{k}\simeq N$ highly
dominate the others with $\avg{k}\simeq N/2^q$.  
We might have a different situation only when $q=\mathop{O}(\ln(N))$, where $\avg{k}=\mathop{O}(1)$,
consistently with what we have shown numerically in the previous paragraphs.
Similar arguments hold for other intermediate values of $\avg{k}$.

Remarkably, this combinatorial argument does not depend on the details of the RGG,
or on the dimension $d$, or the nature of the underlying geometry, which, in principle, can also be non Euclidean.
As soon as we are dealing with a continuous set $\mathcal{D}$ equipped with some distance, 
our combinatorial argument can be equally applied. 

\subsection{An easy-hard-easy transition}
We can better formalize the findings of the previous paragraphs as follows.
Le us consider a rigid ($a=1$) $d$-dimensional Euclidean RGG of total volume 1 and threshold $r\propto N^{-1/d}$ (see Sec. V.A).
For finite $N$ and $N_s$, let $\mathcal{N}(\avg{k})$ be the number of graphs having average connectivity $\avg{k}$ each,
and let $\mathcal{N}=\sum_{\avg{k}} \mathcal{N}(\avg{k})$ be the total number of graphs (for $N_s$ finite the sum runs over a finite set of
possible values of $\avg{k}$). If we introduce the PDF
\begin{eqnarray}
\label{ratioN}
f(\avg{k})=\lim_{N_s\to\infty}\frac{\mathcal{N}(\avg{k})}{\mathcal{N}},
\end{eqnarray}  
our approach shows that, for $N$ large  
\begin{eqnarray}
\label{ratioN1}
&& f(\avg{k}\to b\delta\left(\avg{k}-\Omega_dr^d(N-1)\right)\nonumber \\
&&+(1-b)\delta\left(\avg{k}-(N-1)\right),
\end{eqnarray}  
where $\avg{k}=\Omega_dr^d(N-1)=\mathop{O}(1)$ corresponds to the connectivity of a regular $d$-dimensional lattice, 
$\avg{k}=N-1$ with the connectivity of a fully connected graph, $\delta(x)$ stands for Dirac delta function, and $0<b<1$ is a constant.
Eq. (\ref{ratioN1}) is consequence of the first order phase transition in $\avg{k}$ piloted by $\mu$. 
On the other hand, the combinatorial argument of the previous paragraph leads us to interpret the phase transition as an easy-hard-easy transition.
In fact, when we want to draw graphs with a very low connectivity,  $\avg{k}=\mathop{O}(1)$, or a very high connectivity, $\avg{k}=\mathop{O}(N)$,
we are able to figure out how to locate the $N$ nodes in order to have such connectivities. These two opposite cases are indeed
relatively easy to build, and their extreme ideal limits corresponding to $\avg{k}=\Omega_dr^d(N-1)$, 
and to $\avg{k}=N-1$, are even trivial.
Note that, as the combinatorial argument makes clear, we are still able to build graphs with any desired connectivity $\avg{k}$.
In order to do so, it is in fact enough to split the $N$ nodes into $m=N/(\avg{k}+1)$ groups each containing 
$\avg{k}+1$ nodes which, inside the group, are sufficiently close to each other so that each forms a fully connected graph of $\avg{k}+1$ nodes.
If the ratio $m=N/(\avg{k}+1)$ is not an integer, it is still possible to consider slight variations of this construction in order
to reach the desired average connectivity (for example by introducing some asymmetry among the groups).
However, such solutions are very specific. An algorithm that aims at finding the distribution of nodes producing the target $\avg{k}$, 
and that is based on a uniform sampling
over the node positions, would hardly converge (in a time that scales polynomially with $N$) if $\avg{k}\neq\Omega_dr^d(N-1)$
or $\avg{k}\neq N-1$. In fact, our approach shows that we are in the presence of
an easy-hard-easy transition: the first order phase transition marks the boundary between two easy phases between which, 
in correspondence of the critical point $\mu_c$, there are the hard solutions. 
Furthermore, the above specific solution with $m$ groups has the bad aspect to be in general a disconnected graph (notice that $m<N$).
In fact, in most of practical problems, like in ad hoc mobile-networks, one is interested to find solutions which are also connected graphs.
Our guess is that such hard connected solutions correspond to specific heterogeneous graphs. 
Chosen any point of $\mathcal{D}$ as a reference center, 
in these solutions, the highest connected nodes are located near the center of the domain 
$\mathcal{D}$, whereas the lowest
connected nodes are located near the boundaries of $\mathcal{D}$.
The idea is that, radially, nodes are distributed according to a density that decays exponentially 
with the radial distance with some exponent $\alpha$.
The larger is $\alpha$, the smaller is the region occupied by the nodes and - as a consequence - the lager is $\avg{k}$.
This construction is similar to the construction of hyperbolic RGG where (embedded in a hyperbolic space) nodes are instead uniformly distributed 
(with respect to the hyperbolic metric) \cite{Hyper}. The network so constructed is connected and
can give rise to a wide spectra of cases according to $\gamma$ - the exponent characterizing the degree distribution $p(k)$ - 
which in turn depends on $\alpha$,
but the network could also be non power law.
Let us consider a $d=1$ dimensional RGG with $\mathcal{D}=[0,1]$ and threshold $r$. For a normalized exponential PDF we have
\begin{eqnarray}
\label{Const}
{{\nu}}(x)=\frac{\alpha}{1-e^{-\alpha}}e^{-\alpha x}.
\end{eqnarray}  
It is easy to see that, given $N$ nodes, the expected degree $\bar{k}(x)$ of a node located at $x$ is
\begin{eqnarray}
\label{Const1}
\bar{k}(x)=\frac{Ne^{-\alpha x}}{1-e^{-\alpha}}2\sinh(\alpha r),
\end{eqnarray}  
therefore, for $\avg{k}$ we have
\begin{eqnarray}
\label{Const2}
\avg{k}=\frac{1-e^{-2\alpha}}{(1-e^{-\alpha})^2}N\sinh(\alpha r).
\end{eqnarray}  
Similar expressions hold in any dimension $d$ for a PDF ${{\nu}}(\bm{x})$ that decays exponentially with the radial distance:
${{\nu}}(\bm{x})=b\exp(-\alpha |\bm{x}|)/\Omega_d$, with $b$ normalization constant. 
In particular, it is possible to show that $C_1\exp(-\alpha |\bm{x}|)\leq \bar{k}(\bm{x})\leq C_2\exp(-\alpha |\bm{x}|)$,
where $C_1$ and $C_2$ are two constants.
Eq. (\ref{Const2}) shows that, whatever $r$ is, by properly choosing $\alpha$ we can achieve any desired $\avg{k}$. 
In particular, if we consider the standard choice $r\propto 1/N$, for $\alpha=\mathop{O}(1)$ we have $\avg{k}=\mathop{O}(1)$,
while for $\alpha=\mathop{O}(N)$ we have $\avg{k}=\mathop{O}(N)$, \textit{i.e.}, the regimes corresponding to the two easy phases.
Yet, the number of possible graphs that one can actually build by using this construction strongly depends on $\alpha$,
or on $\gamma$, if the resulting degree distribution is scale free. 
In fact, in \cite{Genio}, Del Genio \textit{et al.} have proved
that scale free graphs with $0<\gamma<2$ are either very rare or do not exist, 
while they exist for $\gamma>2$ and $\gamma<0$. 
Graphs with $\gamma>2$ are networks with $\avg{k}=\mathop{O}(1)$, and are in correspondence to one of the two easy phases
of our satisfaction problem, whereas
graphs with $\gamma<0$ correspond to graph realizations nearly fully connected, and are in 
the other easy phase where $\avg{k}=\mathop{O}(N-1)$.
Graphs with $0<\gamma<2$, if any, 
are instead networks that would be able to 
give rise to $\avg{k}=\mathop{O}(N^\beta)$ with $0<\beta <1$, \textit{i.e.}, graphs in the set 
of the hard solutions of our satisfaction problem (if not empty), but the result of \cite{Genio} forbids their existence.
On the other hand, Eqs. (\ref{Const})-(\ref{Const2}) are exact and define a way to build graphs 
for any $\alpha$ and therefore for any desired $\avg{k}$:
even if the number of such graphs strongly depends on $\alpha$, it is never zero. 
It is possible to show that this conclusion
does not contradict the finding of \cite{Genio}.
In fact, if we indicate with $p(k|\bm{x})$ the conditional probability that a node located at $\bm{x}$ has degree $k$, 
from $p(k)=\int d\bm{x}{{\nu}}(\bm{x})p(k|\bm{x})$ and $p(k|\bm{x})=(\bar{k}(\bm{x}))^k/k!\exp(-\bar{k}(\bm{x}))$ \cite{BogunaHidden},  
by using the asymptotic behavior of the incomplete gamma function, for $d=1$, $r\propto 1/N$ and $N$ large, one has 
\begin{eqnarray}
\label{Const3}
p(k)\sim \frac{\avg{k}^k}{k!}e^{-\avg{k}}, \quad \mathrm{for} \quad \alpha=\mathop{O}(1),
\end{eqnarray}  
\begin{eqnarray}
\label{Const4}
p(k)\sim \mathrm{Constant}, \quad \mathrm{for} \quad \alpha=\mathop{O}(N^\beta),~1\geq \beta>0.
\end{eqnarray}  
Similar expressions can be found as upper bounds for $d>1$.
Eqs. (\ref{Const3}) and (\ref{Const4}) show that, when we are in the sparse regime with $\avg{k}=\mathop{O}(1)$, 
the hard solutions correspond approximately to a Poissonian $p(k)$,
while, when we are in a dense regime, with $\avg{k}=\mathop{O}(N^\beta)$, 
the hard solutions correspond approximately to a uniform $p(k)$, 
where all nodes tend to form nearly fully connected structures. 
The two cases correspond, formally, to $\gamma\to~ +\infty$ and to $\gamma\to~ 0$, respectively,
compatibly with \cite{Genio}.

\section{Conclusions}
By making use of generating function and saddle point techniques, 
we have derived the equations for the typical node distributions ${{\nu}}(h)$ of a generic hidden-variables
model. We have then applied these equations to RGG to face a non trivial satisfaction problem:
Given $N$ nodes, a domain $\mathcal{D}$, and a desired average connectivity $\avg{k}$,
find - if any - the distribution of nodes having support in $\mathcal{D}$ and
average connectivity $\avg{k}$. However, the numerical solutions of these equations for $d=1$ Euclidean RGG shows that the typical ${{\nu}}(h)$, in the thermodynamic limit,
can only be either uniformly distributed or highly condensed, the two regimes
being separated by a first order phase transition characterized by a $\mathop{O}(N)$ jump of $\avg{k}$.
Other intermediate values of $\avg{k}$ correspond in fact to very rare graph realizations.
We have then provided a combinatorial argument to fully explain the mechanism inducing this phase transition in general RGG 
and recognize it as an easy-hard-easy transition triggered by the geometry,
and that the hard and connected solutions correspond to strongly heterogeneous constructions embedded in the geometrical space, 
but these are not necessarily scale free. 
Our result concludes that, in general, ad hoc optimized networks embedded in geometrical spaces can hardly be designed,
unless to rely on very specific constructions.  

In our approach, a crucial mathematical tool has been the use of a chemical potential $\mu$
in order to tune the desired average connectivity $\avg{k}$. Similarly, one could include in the approach
other free parameters in order to control, for example, the average number of triangles,
or other interesting graph metrics. Notice that, once the solution ${{\nu}}(h)$ has been found, one has
access not only to the averages of a graph metric, but also to its higher moments.
It would be interesting, for certain practical problems described via hidden variable models, to investigate
how to better exploit this approach, for example, by requiring that some metrics have also minimal fluctuations,
or by requiring that certain correlations are reproduced, similarly to the analysis performed in \cite{Colomer},
where clustered scale free models are tuned in such a way to reproduce metrics observed in real world networks.

A different urgent question concerns what this phase transition scenario implies for non Euclidean RGG.
We have already mentioned the parallelism with the hyperbolic case \cite{Hyper}.
Although our combinatorial argument suggests that this phase transition is expected to be
present in any RGG defined over a continuous domain,
the issue requires further thoughtful studies.

\begin{acknowledgments}
Work supported 
by CNPq Grant PDS 150934/2013-0.
We thank D. Krioukov for early discussions.
\end{acknowledgments}

\appendix
\section{\label{appA}Absence of typical detailed configurations in hidden-variables models}
In this appendix we show the lack of typical detailed configurations in
the ensemble in which the hidden-variables are not fixed. 
Let be given a hidden variable model via the PDF $\rho(h)$ and the link probability $p(h_i,h_j)$.
The conditional probability to realize a graph $\bm{G}$ with frozen hidden variables $\{h_i\}$ is
\begin{eqnarray*}
&& P(\bm{G}|\{h_i\})=\prod_{i<j}\left[p(h_i,h_j)\right]^{a_{i,j}}\left[1-p(h_i,h_j)\right]^{1-a_{i,j}},
\end{eqnarray*}
so that the PDF to have $\bm{G}$ together with the hidden-variables values $\{h_i\}$ is   
\begin{eqnarray*}
&& P(\bm{G}\cap \{h_i\})=\prod_{i}\rho(h_i)\prod_{i<j}\left[p(h_i,h_j)\right]^{a_{i,j}}\nonumber \\
&&\times \left[1-p(h_i,h_j)\right]^{1-a_{i,j}},
\end{eqnarray*}
while the (unconditioned) probability to have $\bm{G}$ is
\begin{eqnarray*}
&& P(\bm{G})=\int \prod_{i}\rho(h_i)\prod_{i<j}\left[p(h_i,h_j)\right]^{a_{i,j}}\nonumber \\
&&\times \left[1-p(h_i,h_j)\right]^{1-a_{i,j}}.
\end{eqnarray*}
The generating function of these probabilities can be obtained from
\begin{eqnarray}
\label{ZA}
&& Z(\{x_{i,j}\};\{y_i\})\nonumber \\
&&= \int \prod_i dh_i \rho(h_i) e^{y_ih_i}\sum_{\{a_{i,j}\}}P(\bm{G}|\{h_i\})e^{x_{i,j}a_{i,j}},
\end{eqnarray}
where $\{x_{i,j}\}$ and $\{y_i\}$ are link- and node-auxiliary fields, respectively.
Actually, we do not need to use the $y$'s, since they do not have (at least here) an interesting
graph meaning. However, just for completeness, for the moment being we keep the $y$'s general, while we will set
them to 0 later on. 
Let us rewrite $Z$ as (compact notation: $p_{i,j}=p(h_i,h_j)$)
\begin{eqnarray}
\label{Z1}
&& Z(\{x_{i,j}\};\{y_i\})= \int \prod_i dh_i \prod_{i}\rho(h_i) e^{y_ih_i}\nonumber \\
&& \prod_{i<j}~\sum_{a_{i,j}=0,1} e^{a_{i,j}\left(\ln(p_{i,j})+x_{i,j}\right)+\left(1-a_{i,j}\right)\ln\left(1-p_{i,j}\right)}.
\end{eqnarray}
By summing over the $a_{i,j}$, we get
\begin{eqnarray}
\label{Z2}
&& Z(\{x_{i,j}\};\{y_i\})= \int \prod_i dh_i e^{Nf\left(\{x_{i,j}\};\{y_i\};\{h_i\}\right)},
\end{eqnarray}
where
\begin{eqnarray}
\label{f}
&&Nf\left(\{x_{i,j}\};\{y_i\};\{h_i\}\right)=\sum_i\left(y_i h_i+\ln(\rho(h_i))\right)\nonumber \\
&&+\sum_{i<j}\ln
\left(
1-p_{i,j}+p_{i,j}e^{x_{i,j}}
\right).
\end{eqnarray}
For $N$ large we can try to use a ``saddle-point'' technique by solving for the $h$'s the system of the $N$ coupled Eqs. 
\begin{eqnarray}
\label{Df}
\frac{\partial f}{\partial h_i}=0, \quad i=1,\ldots,N.
\end{eqnarray}
From Eq. (\ref{f}) we have
\begin{eqnarray}
\label{Df1}
&&\frac{N\partial f}{\partial h_i}=y_i+\frac{1}{\rho(h_i)}\frac{\partial \rho(h_i)}{\partial h_i}\nonumber \\
&&+\sum_{j: j\neq i}\frac{\left(e^{x_{i,j}}-1\right)}{p_{i,j}\left(e^{x_{i,j}}-1\right)+1}\frac{\partial p_{i,j}}{\partial h_i}.
\end{eqnarray}

Let us specialize now for the following family of hidden-variables models:
\begin{eqnarray}
\label{rhoa}
\rho(h)=ah^{-\gamma},
\end{eqnarray}
\begin{eqnarray}
\label{pa}
p(h_i,h_j)=h_i h_j ~w(h_ih_j),
\end{eqnarray}
where $a$ is a normalization constant, and $w(x)$ is a positive dumping factor such that $x ~ w(x) \leq 1$ for any $x\geq 0$,
which in particular implies $w(x)\to 0$, $w'(x)\to 0$, and $w(x)/w'(x)\to 0$ for $x\to \infty$.
For the model (\ref{rhoa}-\ref{pa}), the derivatives take simple forms:
\begin{eqnarray}
\label{Dp}
\frac{\partial p_{i,j}}{\partial h_i}=\frac{p_{i,j}}{h_i}+\frac{p^2_{i,j}}{h_i}\frac{w'_{i,j}}{w^2_{i,j}}.
\end{eqnarray}
In particular, for the standard case $w(x)=1/(1+x)$, we have
\begin{eqnarray}
\label{Dp}
\frac{\partial p_{i,j}}{\partial h_i}=\frac{p_{i,j}}{h_i}-\frac{p^2_{i,j}}{h_i}.
\end{eqnarray}
For $N$ large and $\gamma>2$,
we can neglect terms in $p^2_{i,j}$ w.r.t. to those in $p_{i,j}$ (this approximation is however not essential here). 
In conclusion, for the model (\ref{rhoa}-\ref{pa}), Eqs. (\ref{Df1}) become
\begin{eqnarray}
\label{Df2}
\frac{N\partial f}{\partial h_i}=y_i-\frac{\gamma}{h_i}+\frac{1}{h_i}
\sum_{j: j\neq i}\frac{p_{i,j}\left(e^{x_{i,j}}-1\right)}{p_{i,j}\left(e^{x_{i,j}}-1\right)+1},
\end{eqnarray}
which leads to the following system of saddle-point Eqs. for the $h$'s
\begin{eqnarray}
\label{Df3}
y_ih_i=\gamma-\sum_{j: j\neq i}\frac{p_{i,j}\left(e^{x_{i,j}}-1\right)}{p_{i,j}\left(e^{x_{i,j}}-1\right)+1}, \quad i=1,\ldots,N.
\end{eqnarray}
For a given set of values of the auxiliary fields $\{x_{i,j}\};\{y_i\}$, Eqs. (\ref{Df3}) can have one or more solutions.
We will indicate a solution of the saddle-point Eqs. (\ref{Df3}) with the superscript $^*$: $\{h_i^*\}$.
Note that $h_i^*=h_i^*\left(\{x_{i,j}\};\{y_i\}\right)$.
If we set $\{y_i=0\}$, Eqs. (\ref{Df3}) reduce to
\begin{eqnarray}
\label{Df4}
\gamma=\sum_{j: j\neq i}\frac{p_{i,j}\left(e^{x_{i,j}}-1\right)}{p_{i,j}\left(e^{x_{i,j}}-1\right)+1}, \quad i=1,\ldots,N.
\end{eqnarray}
However, we immediately see that Eqs. (\ref{Df4}) have no solution for $\{x_{i,j}=0\}$, which is the value of the auxiliary fields
where we have to set up our calculations at the end to get $Z$, and its derivatives (in order to get also the averages).
In other words, in the ensemble where the $h$'s are random variables
there are no typical graphs. 

We can alternatively try to make first the integral over the $h$'s and only later to sum over the $\{a_{i,j}\}$.
We get 
\begin{eqnarray}
\label{ZR}
&& Z(\{x_{i,j}\})\nonumber \\
&&= \sum_{\{a_{i,j}\}}
\int \prod_i dh_i e^{Ng\left(\{x_{i,j}\};\{a_{i,j}\};\{h_i\}\right)},
\end{eqnarray}
where 
\begin{eqnarray}
\label{g}
&&Ng\left(\{x_{i,j}\};\{a_{i,j}\};\{h_i\}\right)=\sum_i\ln(\rho(h_i)) \\
&&+\sum_{i<j}\left[
a_{i,j}\left(\ln(p_{i,j}) +x_{i,j}\right)+
\left(1-a_{i,j}\right)\ln\left(1-p_{i,j}\right)
\right]. \nonumber
\end{eqnarray}
We have
\begin{eqnarray}
\label{Dg2}
\frac{N\partial g}{\partial h_i}=-\frac{\gamma}{h_i}+\frac{1}{h_i}
\sum_{j: j\neq i}\frac{p_{i,j}\left(e^{x_{i,j}}-1\right)}{p_{i,j}\left(e^{x_{i,j}}-1\right)+1},
\end{eqnarray}
which leads to the following system of saddle-point Eqs. for the $h$'s
\begin{eqnarray}
\label{Dg3}
\gamma=\sum_{j: j\neq i}\frac{a_{i,j}-p_{i,j}}{1-p_{i,j}}, \quad i=1,\ldots,N.
\end{eqnarray}
It is not possible to satisfy such saddle-point Eqs..

\section{\label{appB}Percolation in RGG}
Percolation in RGG has been studied for Euclidean RGG's and also rigorously in $d=1$ and $d=2$ 
dimensions \cite{Meester,Penrose}.
It is not the aim of this paper  to analyze in detail the percolation in RGG, however, it should be clear that
the solution ${{\nu}}(h)$ of the saddle point Eqs. (\ref{spq1}) contains crucial information about the percolation problem.
In particular, we can analyze the percolation as follows. 
Let us consider a $d-$dimensional rigid ($a=1$) Euclidean RGG with domain $\mathcal{D}=[0,1)^d$. Here $h$ represents a position vector in 
$[0,1)^d$. Given $r>0$, two nodes are connected if and only if
their euclidean distance is at most $r$.
Given the number of nodes $N$, the initial density $\rho(h)$ (which in turn defines $\mathcal{D}$), 
and the solution ${{\nu}}(h)$ of Eqs. (\ref{spq1}), let us define the following
set in the continuum
\begin{eqnarray}
\label{D1}
\mathcal{D}_1 =\{ h\in \mathcal{D}: N\int_{\mathcal{B}(h,r)}{{\nu}}(h')dh'\geq 1\},
\end{eqnarray}
where $\mathcal{B}(h,r)$ stands for the $d-$ball of radius $r$ centered at $h$.
The set $\mathcal{D}_1$ provides the positions $h$'s where, in average, there is at least one node within the balls $\mathcal{B}(h,r)$'s.
It is then clear that, in the limit in which we can neglect fluctuations of the node positions, the RGG will be percolating if 
$\mathcal{D}_1$ is a connected set in $\mathcal{D}=[0,1)^d$ (in the topological sense).  
Let us consider the standard case where nodes are uniformly sprinkled over $[0,1)^d$, \textit{i.e.}, $\rho(h)=1$. In this case we have also ${{\nu}}(h)=1$,
from which we get that, depending on the value of $r$, we have either $\mathcal{D}_1=\emptyset$, or $\mathcal{D}_1=[0,1)^d$.
Therefore, since $[0,1)^d$ is connected, the RGG is percolating if $r>r_p$, and not percolating if $r<r_p$, where
\begin{eqnarray}
\label{rc}
r_p=\frac{1}{\left(\Omega_d N\right)^{1/d}},
\end{eqnarray}
where $\Omega_d$ is the solid angle in $d$ dimension.
Despite our crude approximation in neglecting the node fluctuations, 
the dependence of $r_p$ on $N$ turns out to be correct (see also \cite{Dall}). 
In fact, in $d=2$ dimension the percolation threshold $r_t$, \textit{i.e.}, the minimal value 
of $r$ where a giant connected component appears, is rigorously known to be $r_t \propto (1/N)^{1/2}$.  
However, the critical value of $r$ above which the RGG is also connected, is rigorously known to be $r_c \propto (\ln(N)/N)^{1/2}>r_t$,
therefore the argument of the above approximation is not fully consistent. However, it has the advantage that it is general and can be applied
to any sprinkle $\rho(h)$, not necessarily uniform or nearly uniform. Clearly, the smaller is the variance of $\rho(x)$,
the better is the approximation.

\end{document}